\documentclass[useAMS,usenatbib]{mn2e}
\usepackage{graphicx}
\title[Spun-up stars and the X-ray luminosity of
  Sgr A*]{Coronal radiation of a cusp of spun-up stars and the X-ray
  luminosity of Sgr A*}

\author[S. Sazonov et al.]{S. Sazonov$^{1,2}$\thanks{E-mail:
sazonov@iki.rssi.ru}, R. Sunyaev$^{2,1}$ and M. Revnivtsev$^{1}$\\
$^{1}$Space Research Institute, Russian Academy of Sciences,
Profsoyuznaya 84/32, 117997 Moscow, Russia\\
$^{2}$Max-Planck-Institut f\"ur Astrophysik,
Karl-Schwarzschild-Str. 1, 85740 Garching bei M\"unchen, Germany
}

\newcommand{\beq}{\begin{equation}}
\newcommand{\eeq}{\end{equation}}
\newcommand{\beqa}{\begin{eqnarray}}
\newcommand{\eeqa}{\end{eqnarray}}

\newcommand{\lesssim}{\hbox{\rlap{\hbox{\lower4pt\hbox{$\sim$}}}\hbox{$<$}}}
\newcommand{\gtrsim}{\mathrel{\hbox{\rlap{\hbox{\lower4pt\hbox{$\sim$}}}\hbox{$>$}}}} 

\newcommand{\NH}{N_{\rm{H}}}

\newcommand{\rs}{R_{\rm{s}}}
\newcommand{\rp}{R_{\rm{p}}}
\newcommand{\ns}{N_{\rm{s}}}
\newcommand{\omegab}{\Omega_{\rm{b}}}
\newcommand{\omegas}{\Delta\Omega_{\rm{s}}}
\newcommand{\omegac}{\Omega_{\rm{c}}}

\newcommand{\vrel}{v_{\rm{rel}}}
\newcommand{\vp}{v_{\rm{p}}}
\newcommand{\sig}{\sigma_{\ast}}

\newcommand{\tmb}{t_{\rm{mb}}}
\newcommand{\lbol}{L_{\rm{bol}}}
\newcommand{\lx}{L_{\rm{X}}}
\newcommand{\lhx}{L_{\rm{HX}}}
\newcommand{\ehx}{\epsilon_{\rm{HX}}}

\newcommand{\tc}{t_{\rm{c}}}
\newcommand{\emab}{\epsilon_{\rm{AB}}}
\newcommand{\dgc}{D_{\rm{GC}}}
\newcommand{\flms}{f_{\rm{LMS}}}
\newcommand{\fsr}{f_{\rm{SR}}}
\newcommand{\msr}{M_{\rm{SR}}}
\newcommand{\mbh}{M_{\rm{BH}}}

\newcommand{\avns}{\langle N_{\rm{s}}\rangle}
\newcommand{\avt}{\langle \Delta t_{\rm{s}}\rangle}

\newcommand{\rd}{R_{\rm{d}}}
\newcommand{\nd}{N_{\rm{d}}}

\newcommand{\ledd}{L_{\rm{Edd}}}
\newcommand{\rb}{R_{\rm{B}}}
\newcommand{\cs}{c_{\rm{s}}}
\newcommand{\mdout}{\dot{M}_{\rm{out}}}
\newcommand{\mdin}{\dot{M}_{\rm{in}}}

\begin{document}

\maketitle

\label{firstpage}

\begin{abstract}
{\sl Chandra} has detected optically thin, thermal X-ray emission with
a size of $\sim 1$~arcsec and luminosity $\sim 10^{33}$~erg~s$^{-1}$
from the direction of the Galactic supermassive black hole (SMBH), 
Sgr~A*. We suggest that a significant or even dominant fraction of 
this signal may be produced by several thousand late-type
main-sequence stars that possibly hide in the central $\sim 0.1$~pc
region of the Galaxy. As a result of tidal spin-ups caused by close
encounters with other stars and stellar remnants, these stars should
be rapidly rotating and hence have hot coronae, emitting copious
amounts of X-ray emission with temperatures $kT\lesssim
$~a~few~keV. The {\sl Chandra} data thus place an interesting upper
limit on the space density of (currently unobservable) low-mass 
main-sequence stars near Sgr~A*. This bound is close to and consistent
with current constraints on the central stellar cusp provided by 
infrared observations. If coronally active stars do provide a significant
fraction of the X-ray luminosity of Sgr~A*, it should be moderately variable on
hourly and daily time scales due to giant flares occurring on
different stars. Another consequence is that the quiescent X-ray luminosity and
accretion rate of the SMBH are yet lower than believed before.
\end{abstract}


\section{Introduction}
\label{s:intro}

Our Galactic nucleus with its supermassive black hole (SMBH) of mass
$\mbh\approx 4\times 10^6$~$M_\odot$ is a subject of intensive studies in radio,
submillimeter, infrared, X-ray and gamma-ray bands (see
\citealt{genetal10} for a recent review). The {\sl Chandra X-Ray
  Observatory}, with its excellent spatial resolution
\citep{weietal02}, has revealed a rich variety of compact and diffuse
X-ray sources in the central square degree of the Milky Way
\citep{munetal04,wanetal06,munetal08,munetal09}. One of the most
interesting findings is the detection  of faint extended X-ray emission
surrounding the radiosource Sgr A* associated with the SMBH 
(\citealt{bagetal03}, hereafter B03). The X-ray source has a radius of
only $\sim 1$'', or $\sim 0.04$~pc for a Galactic Centre (GC) distance
of $\sim 8$~kpc. Its X-ray luminosity is few $10^{33}$~erg~s$^{-1}$,
and the spectrum is consistent with thermal emission from optically thin
$kT\sim 1$--4~keV plasma, absorbed by a column $\NH\sim
10^{23}$~cm$^{-2}$ of neutral gas (B03). This column density is
roughly consistent with the interstellar extinction towards the GC region.

The currently favoured view is that the X-ray emission from the
  central arcsecond around Sgr A* is truly diffuse and associated with
  hot gas flowing towards the central SMBH (\citealt{quataert02}; B03;
  \citealt{yuaetal03,cuaetal06,xuetal06}). According to this 
  scenario, the X-ray luminosity is dominated by thermal gas
  emission near the Bondi radius, $\rb=G\mbh/\cs^2\sim 0.05$~pc
  (where $\cs$ is the gas sound speed, corresponding to a temperature
  of a few $10^7$~K). The {\sl Chandra} observations imply that the 
  outer inflow rate is $\mdout\sim 10^{-5}M_\odot$~yr$^{-1}$. Although
  this rate is very low compared to active galactic nuclei, if the SMBH were
  accreting gas at this rate in radiatively efficient mode its
  bolometric luminosity would be several $\times 10^{40}$~erg~s$^{-1}$,
  i.e. 5 orders of magnitude higher than the total luminosity of Sgr A*
  (predominantly emitted at radio to submillimeter wavelengths). The
  radiative output can be much lower in the case of an
  advection-dominated accretion flow (ADAF). However, there is an
  additional problem that if all of the gas captured 
  at $\rb$ with rate $\mdout$ were flowing towards the SMBH, that
  would contradict an upper limit on the gas density in the inner
  regions imposed by measurements of Faraday rotation in the direction
  of Sgr~A*, which imply that $\mdin\lesssim 10^{-7}M_\odot$~yr~$^{-1}$
  (e.g. \citealt{bowetal03}). Largely to circumvent this problem, radiatively
  inefficient accretion flow (RIAF) models have been proposed for
  Sgr~A*, in which very little mass available at large 
  radii actually accretes onto the black hole 
  (e.g. \citealt{yuaetal03}). To summarise, the question of gas
  accretion onto the SMBH in the GC remains an open and hotly debated issue.

B03 also discussed the alternative possibility of the extended X-ray
emission from Sgr A* being the sum of a large number of point X-ray
sources. However, these authors could not suggest good candidates for being such
objects. Below, we revisit this hypothesis and suggest that combined 
emission of a few thousand late-type main-sequence (MS) stars may
contribute significantly to or even dominate the X-ray emission from
the vicinity of Sgr~A*.

The GC stellar cluster is the densest concentration of stars in the
Milky Way. There are few $10^5 M_\odot$ of stars within $0.25$~pc of
Sgr A*, with the space density of massive ($\gtrsim 10M_\odot$) stars
rising inwards down to the smallest resolvable projected distances of
$\sim 0.005$~pc from the SMBH (see \citealt{alexander05,genetal10} for
reviews). If the space density of presumably much more numerous
late-type MS stars behaves similarly, many of these should be rapidly rotating
as a result of repeated tidal spin-ups caused by close encounters with
other stars and stellar remnants (\citealt{alekum01}, hereafter
AK01). We pursue this idea further by pointing out that fast rotation
in stars with outer convective zones is always associated with strong
coronal activity and greatly enhanced X-ray emission. Hence,  
if there is indeed a high-density cusp of late-type MS stars in the GC, 
which could be verified by near-infrared (NIR) observations with future 
extremely large telescopes, their cumulative coronal radiation can be 
responsible for a significant fraction of the X-ray emission from 
the vicinity of Sgr A*.

\section{The model}

AK01 estimated typical rotational velocities that stars in a GC
stellar cusp can acquire over their lifespans as a result of repeated
tidal interactions. We use their results to calculate the rotational 
evolution of stars. In addition, we take into account a counteracting effect 
-- magnetic braking of stellar rotation.

\subsection{The stellar cusp near the Galactic SMBH}
\label{s:cusp}

The key question and major source of uncertainty for this study
is how many late-type MS stars are present near Sgr A*. Unfortunately, NIR
observations cannot yet provide a direct answer, since Sun-like stars
are too faint and probably too densely clustered in the Galactic
nucleus to be detected and individually resolved by present-day
8--10~meter telescopes.  

High spatial resolution NIR observations, which currently can reliably 
resolve individual stars brighter than $K\sim 17$ in the GC region,
have demonstrated that the space density of such bright stars
increases inwards as $\rho_\ast(r)\propto r^{-\gamma}$, with
$\gamma\approx 1.2$, in the innermost $\sim 0.25$~pc,
whereas the slope steepens to $\gamma\approx 1.75$ at larger distances
(\citealt{schetal07}; the actual slope of the nuclear cluster is
inferred to be yet steeper once the contribution of the Galactic bulge
to stellar counts is taken into account, \citealt{graspi09}). Given
the limit $K\lesssim 
17$ and taking into account the extinction of $A_{K}\sim 3$ towards the GC, this
radial profile represents the combined space density of early-type MS
stars of mass $\gtrsim 5M_\odot$ (i.e. of B and earlier types) and
low-mass giants (see Fig.~2 in \citealt{genetal10}). Furthermore, a
very similar radial profile describes well the surface brightness of
diffuse NIR light in the GC region, i.e. the cumulative emission of
unresolved stars \citep{schetal07,yusetal11}. The diffuse light
is expected to be dominated by medium-  ($\lesssim 5M_\odot$) and low-
($\lesssim 1 M_\odot$) mass MS stars. Hence, one can infer that there
is a central cusp with $\gamma\approx 1.2$ in the space density of
such stars (this issue is further discussed in \S\ref{s:diffuse} below).

Further constraints on the radial profile and composition of the
GC stellar cluster are provided by spectroscopic NIR observations, which
can distinguish early-type MS stars from evolved low-mass stars, but 
are currently limited to $K\lesssim 15.5$. Such studies have demonstrated
that the number density of B dwarfs in the central $\sim 0.5$~pc
monotonically rises towards Sgr~A* with a slope $\gamma\sim 2.5$
\citep{bucetal09,doetal09,baretal10}; in particular, there are several
tens of B stars in the central arcsecond, i.e. within 0.05~pc of
the SMBH -- the famous S-star cluster. In contrast, the surface number
density of red giants is approximately constant within several
arcseconds of Sgr A* \citep{doetal09,bucetal09,baretal10}, which
implies that the radial distribution of evolved low-mass stars is not
steeper than $r^{-1}$.

These observed distributions probably reflect a combination
of effects (see \citealt{alexander05} for a detailed discussion). Two-body  
interactions can establish a relaxed distribution, $\rho_\ast(r)\propto
r^{-(3/2+p_{\rm M})}$, of stars in the GC region on a time scale 
shorter than the Hubble time, with the more massive stars being more
concentrated, $p_{\rm M}\sim 0.25$, than the lighter ones, $p_{\rm
  M}\lesssim 0$ \citep{bahwol77}. In addition, the spatial
distribution of early-type stars can be strongly affected by the
recent history of star formation in the GC region, which is not well
known. Finally, the observed dearth of evolved stars in the
innermost region may be at least partially caused by destruction of giants'
envelopes during collisions with stars and stellar remnants (e.g. 
\citealt{genetal96,alexander99,daletal09}). 

We conclude that existing observations neither strongly support nor reject the
possibility that there is a central cusp with $\gamma\gtrsim 1$ in the
space density of low-mass ($\sim 1 M_\odot$) MS stars. 

\subsubsection{Parametrisation}

Based on the above discussion and following \citet{genetal10}, we assume the
following radial profile of total stellar mass density within 0.25~pc
of Sgr A*:

\beq  
\rho_\ast(r)=A\left(\frac{r}{0.25~\rm{pc}}\right)^{-1.3} M_\odot~{\rm
  pc}^{-3},
\label{eq:rho}
\eeq
where the coefficient $A$ is discussed below. We further
assume that the adopted power-law slope $\gamma=1.3$ pertains to 
the entire MS population in the central 0.25~pc. For completeness, we
also adopt that $\rho_\ast(r)\propto r^{-1.8}$ at $r>0.25$~pc
\citep{genetal10}, but this is of little importance for the present
study, which focuses on the immediate vicinity of Sgr~A*. 

The current best estimate of the coefficient of the mass density profile
(equation~\ref{eq:rho}) is $A\sim 1.5\times 10^6 M_\odot$~pc$^{-3}$, 
with a combined statistical and systematic uncertainty of a factor of
$\sim 2$ \citep{genetal10}. This estimate is based on dynamical
measurements (proper motions and radial velocities of stars) of the
total enclosed mass of the stellar cluster in the central parsec
\citep{trietal08,schetal09} and the assumption that total mass density
follows the number density of bright stars at $r< 1$~pc. It is
important to emphasise that there are no dynamical measurements of the
enclosed mass of the stellar cluster at distances $r\lesssim 0.5$~pc
from Sgr~A*, where the gravitational potential is dominated by the
SMBH, apart from a fairly weak upper limit of $\sim 10^5 M_\odot$ on
any extended mass within $\sim 0.005$~pc \citep{giletal09}.

We further assume a ratio $\flms=0.5$ for the number of MS stars with
masses between 0.4 and 1.5~$M_\odot$ (i.e. early M to late A stars)
per $M_\odot$ of total mass in the GC stellar cusp, and a ratio
$\fsr=0.25$ for the number of stellar remnants (white dwarfs, neutron
stars and black holes) per $M_\odot$ of total mass. These fiducial
numbers are very close to those adopted by AK01, who 
assumed a Salpeter power-law ($\alpha=2.35$) present-day mass function
(PMF) with a low-mass cutoff of $\sim 0.4 M_\odot$. This is a good
approximation for a continuously forming stellar population with the
canonical (such as found in the solar neighbourhood) initial mass
function (IMF), somewhat affected by mass segregation in the potential well of
the  SMBH. As discussed in detail by \citet{loeetal10}, 
this scenario is consistent with the $K$-band luminosity function
measured at $K\lesssim 17$ in the innermost arcsec and also at $>12''$
from Sgr A* \citep{baretal10}, as well as with the ratio of total mass to
diffuse light in the central parsec. However, we stress that there
still remains much uncertainty regarding the star formation history in
the GC region and consequently the PMF of stars and stellar remnants near 
Sgr~A*. In particular, the disc(s) of massive Wolf-Rayet/O stars found
between 1'' and 12'' from Sgr A* demonstrate that a starburst with a
highly unusual, top-heavy IMF occured $6\pm 2$~Myr ago
\citep{pauetal06,baretal10}.

\subsubsection{Constraint from diffuse NIR light measurements}
\label{s:diffuse}

Although, as we stressed above, Sun-like stars cannot yet be directly
observed and counted in the GC region, it is possible to estimate or
at least place upper limits on the number density of low-mass stars
near Sgr~A* by carefully subtracting the contribution of individual
detected stars from NIR images and measuring the surface brightness of 
the remaining diffuse light, presumable made up by weaker, undetectable stars.

Adaptive optics NIR imaging observations performed at 
the ESO VLT \citep{schetal07} suggest that the cumulative surface brightness
of unresolved, $K\gtrsim 17$ stars increases approximately as
$\theta^{-0.2}$ in the central few arcseconds, where $\theta$ is the
projected distance from Sgr~A* in arcsec. The measured enclosed flux density of
diffuse K-band ($\sim 2.3$~$\mu$m) light is 
\beq
F_\nu(<\theta)\approx 0.05\theta^{1.8}~{\rm Jy}.
\eeq

Depending on the (unknown) PMF of stars in the central region, this
cumulative flux can be dominated by unresolved stars of higher or
lower mass. Conservatively assuming that at most half of the total
diffuse light is provided by low-mass ($\sim 0.4$--$1.5 M_\odot$) MS stars,
with $K\sim$~21--23, the above value of the diffuse light flux implies
that there are $\lesssim 3\times 10^4\theta^{1.8}$ Sun-like stars
near Sgr A*, or equivalently their total number within 0.1~pc, 
\beq
N_{\rm LMS}(r<0.1~{\rm pc})\lesssim 6\times 10^4.
\label{eq:n_diff}
\eeq
For our assumed stellar density profile (equation~\ref{eq:rho})
and the adopted value $\flms=0.5$, this corresponds to the following upper
limit:
\beq
A\lesssim 5\times 10^6M_\odot$~pc$^{-3}.
\label{eq:a_diff}
\eeq

Recently, \cite{yusetal11} studied the diffuse NIR light around
Sgr A* using HST observations. They found that the surface
brightness at 1.45--1.9~$\mu$m rises towards Sgr A* as
$\theta^{-0.34\pm 0.04}$ and $\theta^{-0.13\pm 0.04}$ at
$\theta>0.7''$ and $\theta<0.7''$, respectively, and concluded that
the diffuse light is likely dominated by low-mass ($\sim
0.8M_\odot$) MS stars with a total number of $\sim 4\times 10^4$ within
$r=0.1$~pc. This estimate is in good agreement with the upper limit
(equation \ref{eq:n_diff}) derived above from the VLT AO imaging data.

\subsection{X-ray emission from spun-up stars}

\subsubsection{Tidal spin-up}
\label{s:tidal}

AK01 presented a formalism for estimating the tidal spin-up of a
target star of mass $M_\ast$ and radius $R_\ast$ due to a fly-by of 
another star or stellar remnant of mass $m$ with relative velocity $\vrel$ 
and impact parameter $R$. Both the target star and the impactor were assumed 
to be in Keplerian orbits around a SMBH. In addition, AK01 performed 
numerical SPH simulations of close encounters, $R\lesssim$~few~$R_\ast$, 
when the linear formalism becomes inaccurate. In view of the current 
uncertainties associated with the structure and composition of the GC stellar 
cusp and with the efficiency of tidal spin-up, we take a simplistic approach 
that consists of parametrising our model based on the results of AK01.

Our assumptions are as follows. First, for realistic parameters of the
GC stellar cusp, neither MS stars more massive than few $M_\odot$ nor
low-mass giants, with their large moments of inertia, can be
significantly spun up by repeated tidal 
interactions. Furthemore, coronal activity driven by magnetic dynamo is 
limited to stars with $M_\ast\lesssim 1.5 M_\odot$ (i.e. up to late A
stars, \citealt{guedel04}). It is however unclear whether it is sub-solar
mass or 1--1.5$M_\odot$ stars that will dominate in the combined X-ray
emission of the stellar cusp. On one hand, the X-ray luminosity is
a nearly constant fraction of the bolometric luminosity of late-type
stars that rotate in the saturation regime (see equation~\ref{eq:lx}
below), so that $\lx\propto\lbol\propto M_\ast^{4}$ for such 
stars. On the other hand, the efficiency of tidal spin-up decreases
while the efficiency of magnetic braking increases with 
increasing $M_\ast$ (see \S\ref{s:brake} below). In view of the
uncertainties, we made computations assuming that all stars that can
be spun up are Sun-like ($M_\ast=M_\odot$, $R_\ast=R_\odot$ and 
$L_\ast=L_\odot$) and their space density is $\flms\rho_\ast=0.5\rho_\ast$.

Since stars can be tidally spun up not only in mutual
encounters but also due to interactions with stellar remnants, the total
number of impactors is defined by the sum $\flms+\fsr=0.75$ in our
model. As concerns the parameter $\fsr$, white dwarfs, neutron stars
and black holes are all expected to be present in the central
cluster. We simply assume that all stellar remnants have
$\msr=M_\odot$, the same mass as the MS stars.

We next need to define the stellar velocity field. Following AK01,
we assume that the GC stellar cusp has undergone two-body relaxation.
The one-dimensional velocity dispersion of stars is then \citep{bahwol77} 
\beq
\sig(r)=\sqrt{\frac{1}{p_M+5/2}}{\frac{G\mbh}{r}},
\eeq
which is a weak function of stellar mass, with $p_M$ ranging from 0 for 
the least massive stars to $1/4$ for the most massive ones. Since we are
here interested in late-type MS stars, we adopt $p_M=0$, so that
\beq
\sig(r)\approx 
370\left(\frac{r}{0.05~{\rm pc}}\right)^{-1/2}~{\rm km}~{\rm s}^{-1},
\label{eq:sigma}
\eeq
where we have adopted the SMBH mass $\mbh=4\times 10^6M_\odot$ 
\citep{gheetal08,giletal09}. The distribution of relative velocities 
$\vrel$ of interacting pairs of stars (or stars and stellar remnants) 
is approximately the Maxwellian one with a dispersion of 
$\sqrt{6}\sig$ (e.g. AK01).

We now address the question of what kind of encounters are
important. As is detailed in AK01, the efficiency of tidal spin-up
falls rapidly when the periseparation $\rp$ between the impactor and the
target star becomes larger than $\sim$~2--3~$R_\ast$. The spin-up resulting 
from a closer fly-by depends on the impactor's orbit  
in the gravitational potential of the target star and on the mass ratio
of the two components. Crudely, i) the efficiency of tidal spin-up
increases with decreasing periseparation down to $\rp\sim R_\ast$ when
loss of mass during the interaction becomes important, and ii) for
given $\rp\sim 2R_\ast$, the net change $\Delta\Omega$ in the
rotational angular velocity of the target star is inversely
proportional to $\vp$, the relative speed of the components at periastron. 

In line with our simplistic approach, we adopt that 
\beqa
\Delta\Omega &=&
\left\{
\begin{array}{ll}
\omegas(\omegab R_\odot/\vp) & 
{\rm if}~~\rp\le \rs;\\
0 & {\rm if}~~\rp> \rs. 
\end{array}
\right.
\label{eq:domega}
\eeqa
with $\omegas=0.25\omegab$ and $\rs=2.5R_\odot$. 
Here $\omegab\equiv(GM_\ast/R_\ast)^{1/2}$ is 
the star's centrifugal breakup speed, which for solar-type stars
corresponds to an equatorial speed of $\approx 440$~km~s$^{-1}$, or an orbital
period of $\approx 3$~hours ($\approx 200$ times the rotation rate of the
Sun). We further assume that repeated spin-ups ${\bf\omegas}$ have
random orientations with respect to the target star's rotation axis
${\bf\Omega}$ (this, by the way, implies that stars occasionally can 
experience tidal spin-downs).

The periastron speed $\vp$, which enters equation~(\ref{eq:domega}), can be
derived from the relative velocity of the impactor and the target 
star at infinity via the standard relation for hyperbolic orbits, 
\beq
\vp=\vrel\sqrt{1+4\left(\frac{\omegab
    R_\odot}{\vrel}\right)^2\frac{R_\odot}{\rs}} 
\eeq 
(here it is assumed that $\rp=\rs$).  
Therefore, $\vp\ge 2\omegab(R_\odot/\rs)R_\odot=0.8\omegab R_\odot$, 
with the minimum reached at $\vrel=0$, and given equation~(\ref{eq:domega}), 
$\Delta\Omega$ never exceeds $0.3125\omegab$ in our model. 
In fact, since typical relative velocities in the stellar cusp (see
equation~\ref{eq:sigma}) $\vrel=\sqrt{6}\sig\gtrsim \omegab R_\odot$, we
only rarely approach this limit and typically have 
$\Delta\Omega\lesssim 0.1\omegab$.

The rate of encounters experienced by a star located at
distance $r$ from the SMBH and leading to its spin-up can be found
as (\citealt{bintre87}, AK01)
\beqa
\frac{d\ns}{dt}(r)=4\sqrt{\pi}(\flms+\fsr)\rho_\ast(r)\sig(r)\rs^2
\nonumber\\
\times\left[1+\left(\frac{\omegab
    R_\odot}{\sig^2(r)}\right)^2\frac{R_\odot}{\rs}\right], 
\label{eq:rate}
\eeqa
where $\rho_\ast(r)$ and $\sig(r)$ are given by
equations~(\ref{eq:rho}) and (\ref{eq:sigma}), respectively. 
Equation~(\ref{eq:rate}) includes the gravitational focusing term. 

Sufficiently close to the SMBH, also collisions of stars leading to  
their destruction may become important. The characteristic
periseparation for such catastrophic events is $\rd\sim
0.5R_\ast=0.5R_\odot$ (AK01), and the corresponding rate is
\beqa
\frac{d\nd}{dt}(r)=4\sqrt{\pi}(\flms+\fsr)\rho_\ast(r)\sig(r)\rd^2
\nonumber\\
\left[1+\left(\frac{\omegab
    R_\odot}{\sig^2(r)}\right)^2\frac{R_\odot}{\rd}\right]. 
\label{eq:destr_rate}
\eeqa
Therefore, for our adopted parameter values destructive collisions
occur $(\rs/\rd)^2\lesssim 25$ (the exact value weakly depends 
on the distance from the SMBH through the gravitional focusing term) times 
less frequently than spin-up encounters. 

Our choice of values for the parameters characterising tidal spin-up
is further discussed in \S\ref{s:test}, where we compare the results
of our test simulation with the results of a simulation with a
similar set-up by AK01.

\subsubsection{Magnetic braking}
\label{s:brake}

A single spinning star cannot sustain its rotation indefinitely,
because it loses angular momentum with the stellar wind via magnetic
braking. Following usual practice (see \S13 in \citealt{schzwa00}), we
parametrise this effect as follows:
\beqa
\frac{d\Omega}{dt} &=&
\left\{
\begin{array}{ll}
-\tmb^{-1}\Omega & 
{\rm if}~~\Omega\ge \omegac;\\
-\tmb^{-1}\left(\frac{\Omega}{\omegac}\right)^2\Omega & 
{\rm if}~~\Omega< \omegac. 
\end{array}
\right.
\label{eq:brake}
\eeqa
Here, $\omegac$ is the angular velocity at which coronal activity
reaches saturation (see \S\ref{s:xray} below). We adopt
$\omegac=0.1\omegab$, which is  approximately equivalent to the
condition ${\rm Ro}'\approx 0.1$ for the modified Rossby number (which
is approximately the ratio of the stellar period and the
characteristic convective turnover time) often used as an activity
saturation criterion for late-type stars (see \citealt{guedel04} for
review). In the standard situation of a star-forming region, the
characteristic time $\tmb$ in equation~(\ref{eq:brake}) has the
meaning of duration of stellar ``youth'', a period when stellar
activity is at its maximum. Observations of stellar clusters of
different ages and stars in the field suggest that
$\tmb\sim$~50--100~Myr for solar-type stars and that this time
increases with decreasing $M_\ast$, perhaps up to $\sim 200$~Myr for
K/M dwarfs \citep{guedel04}. At age $t\gg\tmb$, the decelaration of a
star follows the Skumanich law \citep{skumanich72}, $\Omega(t)\propto
t^{-1/2}$, in agreement with equation~(\ref{eq:brake}).

Since the efficiency of magnetic braking is determined by stellar mass 
and rotation rate, the empirical relations for $d\Omega/dt$ pertaining
to stars in the solar neighbourhood should be equally
applicable to the GC stellar cusp, with the only difference being that
slow-down intervals, governed by equation~(\ref{eq:brake}), can now
alternate with abrupt tidal spin-ups.

\subsubsection{Rotation--luminosity relation}
\label{s:xray}

The X-ray luminosity of a coronally active star depends on its
rotation rate approximately as \citep{guedel04}
\beqa
\frac{\lx}{\lbol} &=&
\left\{
\begin{array}{ll}
10^{-3} & {\rm if}~~\Omega\ge \omegac;\\
10^{-3}(\Omega/\omegac)^2 & {\rm if}~~\Omega< \omegac. 
\end{array}
\right.
\label{eq:lx}
\eeqa

Most of the emission is at energies below 1~keV. However, in the GC
region, with its large extinction, we are more interested in hard
X-ray emission at energies above 2~keV. There is a well-known trend of
spectral hardening with increasing coronal X-ray luminosity. Following
\cite{sazetal06} and in view of equation~(\ref{eq:lx}), we adopt for
the luminosity of solar-type stars in the 2--10~keV energy band
\beq
\lhx=0.25 \left(\frac{\lx}{10^{-3}\lbol}\right)^{0.25}\lx.
\label{eq:lhx}
\eeq

It is important to note that the above formulae, used in our
computations, are, strictly speaking, only appropriate for the
``quiescent'' state of coronally active stars. In reality, as we
discuss in detail below (in \S\ref{s:var}), coronally active stars
occasionally experience giant flares during which their X-ray
luminosity increases by several orders of magnitude and can become
comparable to the bolometric luminosity of the star; moreover, the
emission becomes harder. If several such giant flares occur
simultaneously in the GC stellar cluster at any given time, the
cumulative X-ray (2--10~keV) luminosity of the entire cluster can be
larger by tens of per cent or even by a factor of few than would be expected
based on equations~(\ref{eq:lx}) and (\ref{eq:lhx}). 

\subsection{Contribution from active binaries}
\label{s:ab}

Almost certainly, rapidly rotating single stars are not the
only kind of X-ray sources present in the GC stellar cusp. Among other
plausible contributors to the cumulative X-ray emission are coronally active
stellar binaries (ABs) of RS CVn, BY Dra and Algol types. In ABs, we
deal with a star (in fact two stars) that rapidly rotates due to 
synchronisation with the orbital motion in a close binary. 

We have recently demonstrated that ABs together with cataclysmic
variables (CVs) produce the bulk of the Galactic ridge X-ray emission
(GRXE), the apparently diffuse background observed all over the Milky Way
including the GC region \citep{revetal06,revetal07,revetal09}. The
combined emissivity of ABs and CVs per unit stellar mass in the
2--10~keV energy band is $\sim 3\times
10^{27}$~erg~s$^{-1}$~$M_\odot^{-1}$, of which $\sim 1.5\times
10^{27}$~erg~s$^{-1}$~$M_\odot^{-1}$ is provided by ABs with 
$\lhx<10^{31}$~erg~s$^{-1}$ \citep{sazetal06}. Since the space density
of CVs and ABs with $\lhx>10^{31}$~erg~s$^{-1}$ is only $\sim 10^{-5}
M_\odot^{-1}$ \citep{sazetal06}, we expect at most a few such systems to be
present within 0.25~pc of the SMBH and none within the innermost
0.05~pc, given the mass profile (equation~\ref{eq:rho}). If we assume
that the relative numbers of ABs and solar-type stars are the same as 
in the Solar vicinity and take into account that our (implicitly) 
adopted stellar mass function has a low-mass cutoff of $\sim 0.4 M_\odot$ (see 
\S\ref{s:cusp}), we can expect ABs with $\lhx<10^{31}$~erg~s$^{-1}$ to
produce $\emab\sim 2.5\times 10^{27}$~erg~s$^{-1}$~$M_\odot^{-1}$ in
the GC stellar cusp. The contribution of ABs can be estimated by
multiplying this fractional emissivity by the total mass density $\rho_\ast$. 

In reality, there might be fewer ABs per unit stellar mass near
Sgr A* than in the Solar vicinity because binary stars can be destroyed by
encounters with single stars in dense stellar environments. 
Although luminous ($\lhx\sim 10^{30}$--$10^{31}$~erg~s$^{-1}$) RS CVn systems 
have orbital periods $\sim 1$--10~days and separations $\sim$ a few 
$10^{-2}$~a.e. \citep{stretal93}, even such tight systems can 
be evaporated on a time scale of $10^8$--$10^{10}$~years within 0.25~pc 
of Sgr A* \citep{perets09}, while one should take into account the fact that 
one or both components of RS CVn binaries are typically in their
subgiant phase of evolution, i.e. such systems are usually old. 
Therefore, our adopted fractional X-ray emissivity of ABs should in fact 
be considered an upper limit on the actual conribution of such systems.

\subsection{Final set-up}

\begin{table}
\caption{Model parameters.
\label{tab:model}
}
\smallskip

\begin{tabular}{cccc}
\hline
\multicolumn{1}{c}{Symbol,} &
\multicolumn{1}{c}{Explanation} &
\multicolumn{1}{c}{Adopted value or} \\
\multicolumn{1}{c}{acronym} &
\multicolumn{1}{c}{} &
\multicolumn{1}{c}{allowed range} \\
\hline
A & Mass density coefficient & $\lesssim 5\times 10^6M_\odot~{\rm pc}^{-3}$ \\  
$\flms$ & Fraction of low-mass stars & 0.5 per $M_\odot$\\
$\fsr$ & Fraction of remnants & 0.25 per $M_\odot$\\
$M_\ast$ & Star's mass & $1M_\odot$\\
$\msr$ & Remnant's mass & $1 M_\odot$\\
$\rs$ & Spin-up collision radius & $2.5R_\odot$ \\
$\rd$ & Destructive collision radius & $0.5R_\odot$ \\
$\omegas$ & Spin-up velocity & $0.25\omegab$ \\
$\omegac$ & Saturation velocity & $0.1\omegab$ \\
$\Omega_0$ & Initial velocity & $0.1\omegab$ \\
$\tmb$ & Magnetic braking time & $\sim 50$--200~Myr \\
$\tc$ & Cluster's lifetime & 10~Gyr\\
$\emab$ & Emissivity of ABs & $\lesssim 2.5\times
10^{27}$~erg~s$^{-1}$~$M_\odot^{-1}$\\ 
$\mbh$ & SMBH mass & $4\times 10^6 M_\odot$\\
$\dgc$ & Sgr A* distance & 8~kpc\\
\hline
\end{tabular}
\end{table}

There are a few additional parameters in our model. One is the
lifetime of the GC stellar cusp, $\tc$. We adopt
$\tc=10^{10}$~yr and assume that stars are formed at a constant rate 
over this time span. Another one is the rotational velocity of newly born
stars, $\Omega_0$. We adopt $\Omega_0=0.1\omegab$, which implies that
stars begin their lives with marginally saturated 
coronal activity. Finally, we assume the distance to the GC to be $\dgc=8$~kpc
\citep{gheetal08,giletal09}. 

The parameters of the model are summarised in
Table~\ref{tab:model}. The key parameters, i.e. those whose
uncertainty strongly affects the predicted X-ray luminosity of the GC 
stellar cusp, are the mass density coefficient $A$ and, to
a lesser degree, the characteristic magnetic braking time $\tmb$. We
allowed their values to vary within the indicated ranges, consistent
with existing observations of the GC region and nearby stars,
respectively. Also, as was discussed in \S\ref{s:ab}, the contribution of 
active binaries, $\emab$, can vary from essentially zero to the value quoted in
Table~\ref{tab:model}. The values of the other parameters are fixed in the 
model. Most of these can be considered to be well determined: first of
all $\mbh$, $\dgc$, $\omegac$ and $\Omega_0$, and to a lesser degree
$\rs$, $\rd$ and $\omegas$. The lifetime of the stellar cusp, $\tc$, has 
very little influence on the results of the simulations. The relative  
fractions and typical masses of low-mass stars and stellar remnants
($\flms$, $\fsr$, $M_\ast$, $\msr$) may actually be significantly different
from our adopted values, but this uncertainty is largely absorbed in
the uncertainty on the mass density coefficient $A$. 

We carried out computations by combining numerical integration over
the radial mass profile (equation~\ref{eq:rho}) with a 
simple Monte-Carlo approach for individual stars within a given
radial bin. For each star, first its age, $t_0$, is drawn randomly
between 0 and $\tc$. Then the fate of the star is followed from
$-t_0$ until $t=0$ (present day). During each small interval $dt$, the
star can either spin down by magnetic braking or experience a tidal
impact leading to its spin-up. In the latter case, the star's
angular velocity increases as
${\bf\Omega}(t+dt)={\bf\Omega}(t)+{\bf\Delta\Omega}$, where the
direction of ${\bf\Delta\Omega}$ is drawn randomly whereas its  
amplitude is calculated according to a relative velocity $\vrel$ that
is drawn randomly from the Maxwellian distribution with a dispersion equal to
$\sqrt{6}\sig(r)$ (recall that tidal spin-up is most efficient
for low-velocity encounters). At $t=0$, the X-ray luminosity of the 
star is calculated according to its final angular velocity, $\Omega(0)$.

\section{Results}
\label{s:results}

Below, we first discuss (\S\ref{s:properties}) the computed
intrinsic properties of cusps of spun-up, X-ray luminous stars, such
as the number of spin-up events per star, distribution of 
rotational velocities of spun-up stars and X-ray luminosity density as
a function of distance from the SMBH. We then compare
(\S\ref{s:brightness}) the computed X-ray surface brightness profiles
with an actual X-ray image of Sgr A*, to which end we analyse archival
{\sl Chandra} data. We then present (\S\ref{s:scale}) approximate
fitting relations that allow one to predict the X-ray luminosity and
radial profile of a GC stellar cusp for a given mass density coefficient $A$. 
Finally, in \S\ref{s:spec} we analyse the {\sl Chandra} spectrum of
Sgr A* and discuss its consistency with the expected X-ray spectrum of 
an ensemble of coronally active stars. 

\subsection{Properties of the cusp of spun-up stars} 
\label{s:properties}

\begin{figure}
\centering
\includegraphics[bb=20 170 530 700, width=\columnwidth]{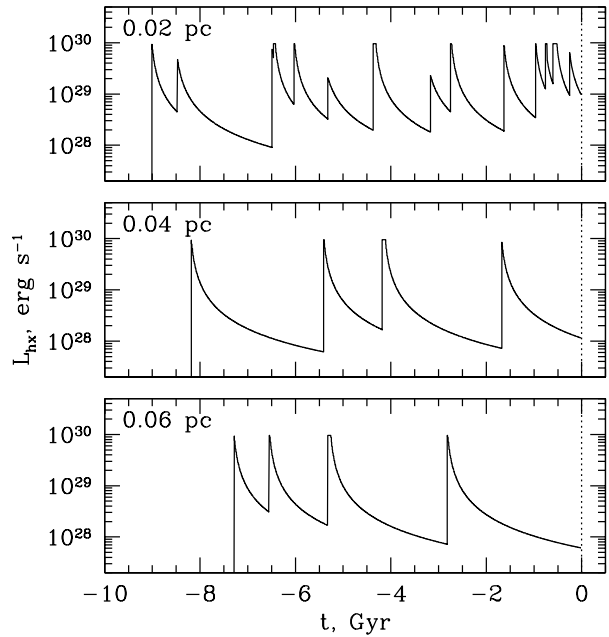}
\caption{Examples of X-ray light curves for stars located at different
  distances from the SMBH, for $A=4\times 10^6 M_\odot$~pc$^{-3}$,
  $\tmb=100$~Myr and other parameter values as given in Table~\ref{tab:model}.} 
\label{fig:history}
\end{figure}

Figure~\ref{fig:history} shows examples of synthetic X-ray light curves of stars
located at different distances from the SMBH. The parameter values are
as quoted in Table~\ref{tab:model}, in particular $A=4\times 10^6
M_\odot$~pc$^{-3}$ and $\tmb=100$~Myr. We see sharp spin-ups caused by
tidal interactions and subsequent long periods of decaying coronal
activity. Outbursts have very short rise times\footnote{The minimal
  possible time scale is determined by the characteristic fly-by time $\sim
2\rs/\vrel\sim 1$~hour, however, the adjustment of the stellar magnetic field
and outer layers to the new rotation rate will probably proceed on the
magnetic dynamo cycle time scale, which is $\sim 1$--10~years  
\citep{saabra99}.} and characteristic durations (FWHM)
$\sim\tmb\sim 100$~Myr. Since our model essentially ignores
inefficient, distant collisions, namely those with
$\rp>\rs=2.5R_\odot$ (see equation~\ref{eq:domega}), it typically takes
just one spin-up episode to bring a star into the 
regime of coronal saturation ($\Omega\sim\omegac$, and hence $\lhx\sim
10^{30}$~erg~s$^{-1}$). The cumulative X-ray emission of the stellar cusp is 
dominated by those stars that are currently caught shortly after the
latest spike in their light curve. 

\begin{figure}
\centering
\includegraphics[bb=20 160 580 720, width=\columnwidth]{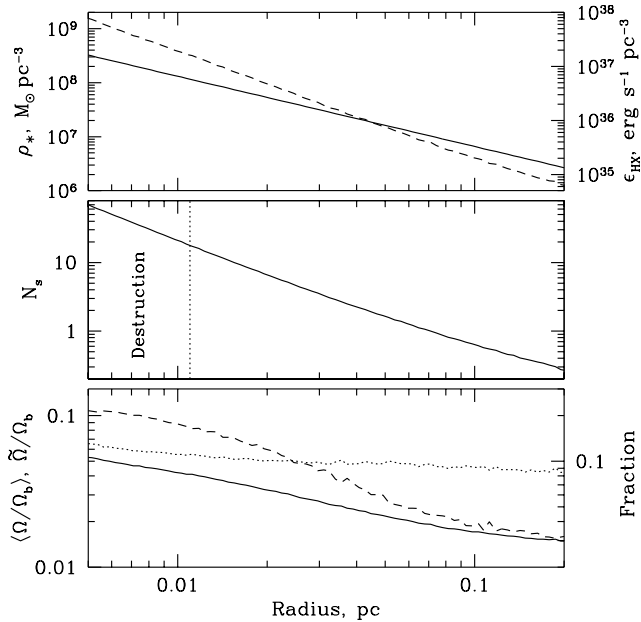}
\caption{Radial profiles of average stellar properties in the GC cusp 
 for $A=4\times 10^6 M_\odot$~pc$^{-3}$, $\tmb=100$~Myr
  and other parameter values as given in Table~\ref{tab:model}. {\sl
    Top panel:} Mass density (solid line and left axis) and 2--10~keV  
  luminosity density (dashed line and right axis). {\sl Middle panel:}
  Number of spin-ups per star. In the region to the left of the
  vertical dotted line, stars are likely to be destroyed by a
  catastrophic collision. {\sl Bottom panel:} Angular rotational
  velocity (left axis): its average value
  $\langle\Omega/\omegab\rangle$ (solid line) and a characteristic 
  value $\tilde{\Omega}/\omegab$ such that stars with 
  $\Omega\ge\tilde{\Omega}$ produce 50\% of the total (excluding the
  contribution of ABs) 2--10~keV emission at a given radius (dotted
  line). Also shown is the corresponding relative fraction of such
  rapid rotators (dashed line and right axis).} 
\label{fig:source}
\end{figure}

Figure~\ref{fig:source} shows the radial profiles of key average properties 
of the stellar cusp. The top panel shows mass density, $\rho_\ast$,
and X-ray (2--10~keV) luminosity density, $\ehx$, as functions of
distance from Sgr A*. The dependence $\ehx(r)$ can be approximately
described by equation~(\ref{eq:ehx_fit}) below. The middle panel of
Fig.~\ref{fig:source} shows the average number of 
tidal spin-up events experienced by stars as a function of
radius. Finally, the bottom panel shows the radial profile of angular
rotational velocity. Specifically, we present the average value
$\langle\Omega/\omegab\rangle$ and a characteristic value 
$\tilde{\Omega}/\omegab$ such that stars with $\Omega\ge\tilde{\Omega}$
produce half of the total X-ray emission at a given radius; also shown
is the relative fraction of such rapid rotators.

We see that stars experience on average $\avns\sim 70$, $\sim 20$, 
$\sim 7$ and $\sim 1.5$ spin-up events over their lifetimes at 0.005, 0.01, 
0.02 and 0.05~pc from the SMBH, respectively. Outside of the innermost
0.006~pc region, the average time between impacts is longer than the
characteristic magnetic braking time: $\avt\sim 0.5\tc/\avns>\tmb$. As
a result, the majority of stars have managed to spin down after their
last tidal interaction. However, a noticeable fraction of stars had their 
last spin-up not more than a few $\tmb$ ago and it is these stars that 
produce the bulk of the X-ray emission in the GC stellar cusp. For example, 
half of the total emission at 0.01~pc from the SMBH is produced by
18\% of stars, namely those having $\Omega\gtrsim 0.055\omegab$, half
of the emission at 0.02~pc by 12\% of stars, with $\Omega\gtrsim
0.05\omegab$, and at 0.05~pc by 6\%, also with $\Omega\gtrsim
0.05\omegab$.

Within 0.011~pc of the SMBH, stars are likely to be destroyed by a
catastrophic collision (see the middle panel of Fig.~\ref{fig:source}
and \S\ref{s:tidal} above). This will probably lead to a significant
depletion of the stellar cusp. In addition, by providing gas for
accretion onto the central SMBH, such collisions can cause strong
outbursts of Sgr~A* (see \S\ref{s:gas} below). 

\subsubsection{Comparison with AK01}
\label{s:test}

We  made a special computation for the case of 10-Gyr old stars that 
begin their lives with no rotation, are subsequently spun up by repeated 
tidal interactions and do not lose angular momentum via magnetic braking, 
for a stellar cusp with $A=3\times 10^{6}M_\odot$~pc$^{-3}$. These
conditions are nearly identical to the scenario considered by AK01. 

The stars acquired high rotational velocities over their
lifespans, $\langle\Omega\rangle\approx 0.26\omegab$, $\approx
0.17\omegab$ and $\approx 0.10\omegab$ at 0.02, 0.05 and 0.1~pc from
the SMBH, respectively. These values are close to but somewhat lower than the
corresponding numbers in AK01 (see their Fig.~7), $\sim 0.3\omegab$,
$\sim 0.2\omegab$ and $\sim 0.15\omegab$, respectively. This
indicates that our choice of values for the $\rs$ and $\omegas$ parameters,
which define the efficiency of tidal spin-up in our simplified treatment,
is reasonable and not overly optimistic.

\subsection{X-ray surface brightness profile}
\label{s:brightness}

\begin{figure}
\centering
\includegraphics[bb=0 0 560 560, width=\columnwidth]{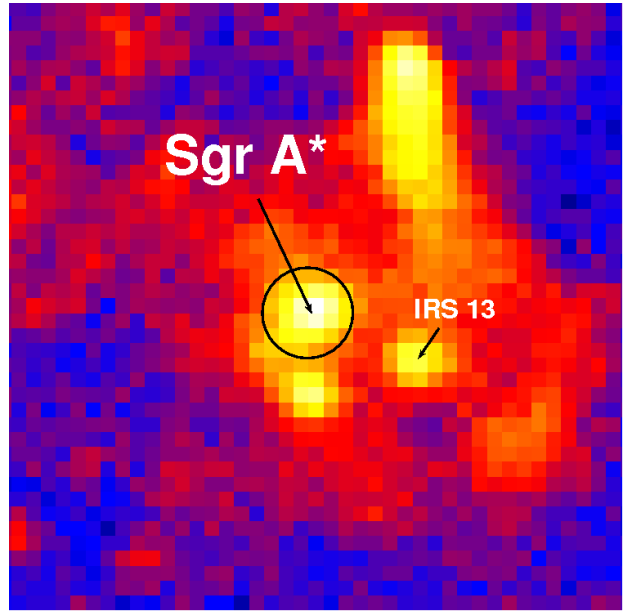}
\caption{{\sl Chandra} image (north is up and east is left) of the
  central $\sim 20''\times 20''$ of the Galaxy in the 2--8~keV energy band.  
The angular size of each pixel is $0.492''\times 0.492''$. The positions 
of the radiosource Sgr A* and the star cluster IRS~13 are marked. Flares 
of Sgr A* have been filtered out. The circle with a radius of 3~pixels drawn 
around Sgr~A* denotes the region used for X-ray spectral analysis.}
\label{fig:image}
\end{figure}

We now address the X-ray surface brightness profile of Sgr A*. Since the
observations reported by B03, the Sgr A* field has been
re-observed with {\sl Chandra} many times. The total effective
exposure accumulated is $\sim 1.2$~Ms, which represents a factor of 
$\sim 25$ increase with respect to B03. We used these archival data to 
measure the X-ray surface brightness around Sgr A*. 

Since apart from the quiescent emission studied in this paper, Sgr A* also 
exhibits X-ray flares \citep{poretal08}, which likely originate in the inner 
regions of an accretion flow onto the SMBH (as suggested by 
multiwavelength and in particular radio interferometric observations,
\citealt{yusetal09}; see however our discussion in \S\ref{s:var}), we made 
an attempt to filter out such flares from the {\sl Chandra} data. Specifically, 
we excluded observations in which increases by a factor of 4 or more in the 
X-ray flux from Sgr A* integrated over 100~s intervals were detected. The 
filtered data set has a total effective exposure of 620~ks. The resulting 
image of the central $\sim 20''\times 20''$ region of the Galaxy obtained 
with the ACIS-I detector is shown in Fig.~\ref{fig:image}. 

\begin{figure}
\centering
\includegraphics[bb=20 160 580 680, width=\columnwidth]{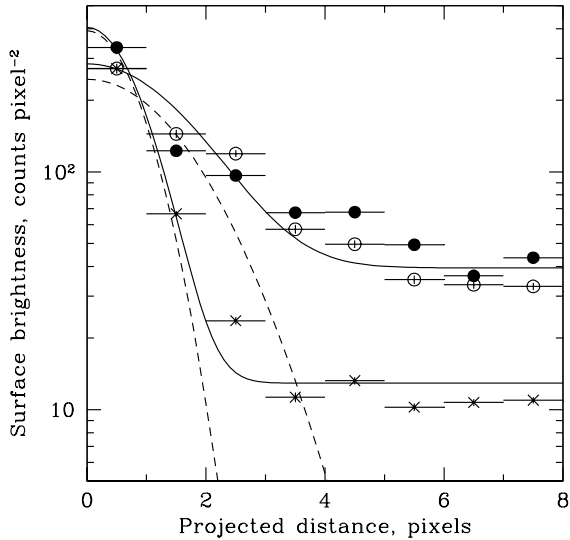}
\caption{Radial surface brightness profiles of the X-ray emission to
  the north and to the south of Sgr~A* (open and filled circles,
  respectively), in comparison with the radial profile of the nearby source
  CXOGC~J174540.9$-$290014 (crosses) normalised to 
  match the central peak of the northern profile of Sgr~A*. Also shown
  are the fits of the northern profile of Sgr~A* and the profile of
  CXOGC~J174540.9$-$290014 by a sum of a constant and a Gaussian with
  $\sigma=1.45$ and 0.75~pixels, respectively (solid lines). The
  Gaussian components are shown separately with the dashed lines.}
\label{fig:xray_profiles}
\end{figure}

Even though we display only the innermost part of the GC region  in
Fig.~\ref{fig:image}, the X-ray image still exhibits a lot of
detail apart from the source at Sgr~A*. In particular, there 
are two bright features within 5'' of Sgr A*, including the 
complex IRS~13 of hot, massive stars (see \citealt{genetal10} for 
a review). Both of these features are located to the south of Sgr~A*. 
We therefore constructed two surface brightness profiles by integrating 
{\sl Chandra} counts in one pixel-wide semi-annuli to the north and to the 
south of Sgr A* (Fig.~\ref{fig:xray_profiles}). For comparison we show
the radial profile of the presumably point-like, bright source
CXOGC~J174540.9$-$290014 located $\sim 20''$ from Sgr~A* (and hence
not shown in Fig.~\ref{fig:image}), which was also discussed by
B03. This profile can be fitted (taking into account 
the finite width of the pixels) reasonably well by a sum of a constant
and a Gaussian with $\sigma\approx 0.37''$. The latter can be used as an
approximation of the point-spread function (PSF) for our mosaic {\sl Chandra} 
image of the Sgr~A* region. Fitting the northern radial profile
of Sgr~A* with the same model gives $\sigma\approx 0.7''$, whereas the
southern profile is significantly non-Gaussian. We can therefore estimate the
intrinsic (i.e. PSF subtracted) size of the central source at $\approx
0.6''$. Our estimates of the PSF width and the intrinsic size of the emission
centred at Sgr~A* are in good agreement with the original results of B03.
  
\begin{figure}
\centering
\includegraphics[bb=20 160 580 680, width=\columnwidth]{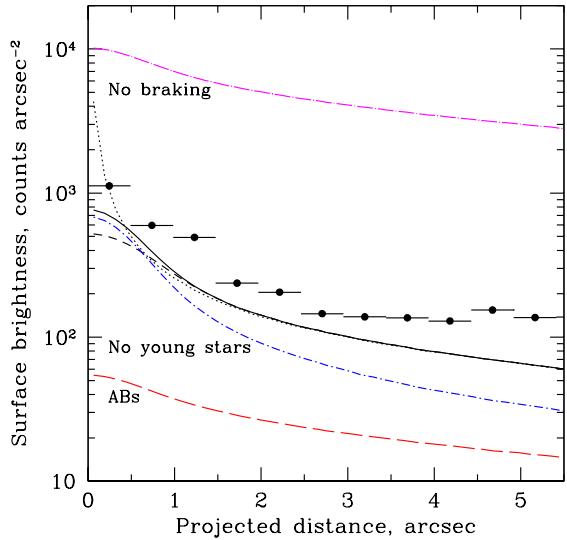}
\caption{Surface brightness profile of the extended X-ray 
  emission around Sgr A*. The data points are {\sl Chandra} 2--8~keV
  measurements to the north of Sgr A*. The dotted line is a model with
  $A=4\times 10^6 M_\odot$~pc$^{-3}$, $\tmb=100$~Myr and other 
  parameter values as given in Table~\ref{tab:model}. There is a $\sim
  20$\% systematic uncertainty in the normalisation of the model,
  associated with the conversion of the intrinsic (unabsorbed) flux to
  {\sl Chandra} counts. The solid line is the same model convolved
  with a Gaussian with $\sigma=0.37$'', approximating the {\sl
    Chandra} PSF. The short-dashed line is the model assuming that there are
  no stars in the central ``destruction'' zone (see
  Fig.~\ref{fig:source}). The magenta dot-long-dashed line illustrates
  what would be if the stars experienced no magnetic braking. The blue
  dot-short-dashed line is a modification of the model, in which all
  stars are born simultaneously 10~Gyr ago, so that there is no
  contribution from young (and therefore coronally active) stars. Finally, the
  maximal expected contribution of active binaries is shown with 
  the red long-dashed line.} 
\label{fig:profile}
\end{figure}

Figure~\ref{fig:profile} compares the {\sl Chandra} radial profile
measured to the north of Sgr~A* with the X-ray surface brightness
profile expected for a stellar cusp with $A=4\times 10^6 M_\odot$~pc$^{-3}$,
$\tmb=100$~Myr and other parameter values as given in
Table~\ref{tab:model}, i.e. the same model as in
Fig.~\ref{fig:source}. We have converted all modelled profiles to {\sl
  Chandra} counts using the spectral model described below in 
\S\ref{s:spec}. The dotted line is the model without taking into
account the angular resolution of the telescope, while the solid line
shows the model convolved with a Gaussian with $\sigma=0.37$'' to
approximate the {\sl Chandra} PSF. We see that the unconvolved model surface
brightness in the innermost region is approximately inversely
proportional to the projected distance (see
equation~\ref{eq:brightness_fit} below). Finally, the dashed line
shows the same model but assuming that the  
central ``destruction'' zone (see Fig.~\ref{fig:source}) is devoid of
stars. Since complete depletion of the innermost cusp by collisions is
unlikely, the solid and dashed curves may be regarded as an upper and
lower limits on the expected emission, respectively.
      
The model reproduces the innermost part of the 
{\sl Chandra} surface brightness profile fairly well, underpredicting
the measured X-ray flux from the central arcsecond by $\sim 25$\%, which
is comparable with the uncertainty in the conversion of the intrinsic
(i.e. unabsorbed) source flux to {\sl Chandra} counts (see
Table~\ref{tab:fits}). For comparison, we show with the magenta
dot-long-dashed line in Fig.~\ref{fig:profile} the same model,
but assuming that the stars experience no magnetic braking. The
resulting X-ray emission is more than an order of magnitude stronger
than is observed, because all stars now spin with $\Omega>0.1\omegab$,
as a result of the rotation received at birth and in tidal
interactions, and produce coronal emisssion with maximal possible efficiency. 

The effect of X-ray emission induced by tidal spin-ups is
limited to $\sim 1.5''$ ($\sim 0.06$~pc) from Sgr
A*. However, our model is also able to account for a significant fraction of
the apparently diffuse X-ray emission detected by {\sl Chandra} further 
away from Sgr A*. In our model, which assumes continuous star formation over 
$\tc=10$~Gyr, this additional emission is partly associated with late-type 
MS stars that were born at most several magnetic braking times (i.e. a few
$10^8$~yr) ago and hence continue to rotate fast and produce strong
X-ray emission. Note that the region between 1'' and 12''
from Sgr~A*, where a starburst apparently occured just $\sim 6$~Myr ago 
\citep{pauetal06}, is expected to contain a large number of young, strongly 
X-ray emitting stellar objects that may account for a significant fraction 
of the apparently diffuse X-ray emission \citep{naysun05}. 

An additional significant contribution can be provided
by active binaries (see the red long-dashed line line in
Fig.~\ref{fig:profile}) if such systems are present and have not been
evaporated  in the GC central cusp (see the discussion in \S\ref{s:ab}).  
This combination of young stars and ABs resembles the situation 
in the solar neighbourhood, where young single stars, ABs and catacalysmic 
variables all provide similar contributions to the cumulative X-ray 
emission \citep{sazetal06}. 

Importantly, the X-ray flux from the innermost $\sim 1.5''$ 
region is essentially determined by the total mass of late-type MS stars
in the central cusp and is almost insensitive to the history of star
formation preceeding the present epoch. This is illustrated in
Fig.~\ref{fig:profile} with the blue dot-short-dashed line, which is
the result of a computation for the same model as before but assuming
that all stars were born at the same time 10~Gyr ago. The flux from
the vicinity of Sgr~A* is nearly the same as in the case of continuous star
formation, but the emission outside $\sim 1.5''$ from Sgr~A* has
decreased due to the disappearance of young stars. 

\begin{figure}
\centering
\includegraphics[bb=20 160 580 680,
  width=\columnwidth]{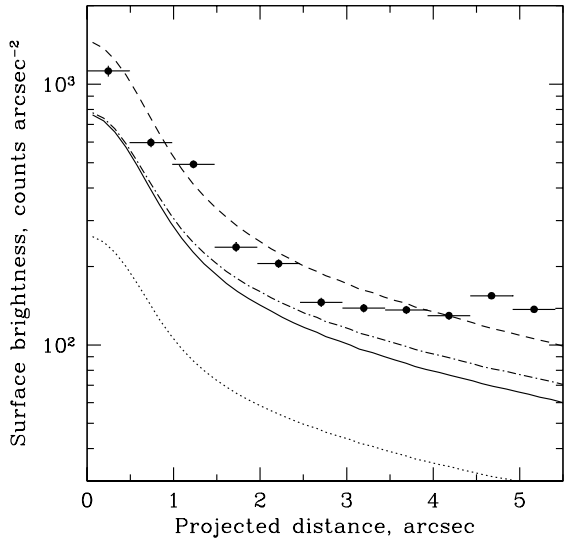}  
\caption{As Fig.~\ref{fig:profile}, but for a set of different models. The
  dotted, solid and dashed lines are for $\tmb=100$~Myr and
  $A=2$, 4 and $6\times 10^6 M_\odot$~pc$^{-3}$, respectively 
  (collisional destruction of stars in the vicinity of the SMBH is
  not taken into account). The dash-dotted line is for $\tmb=200$~Myr
  and $A=3\times 10^6 M_\odot$~pc$^{-3}$.} 
\label{fig:profile_param}
\end{figure}

The uncertainties associated with some of our model parameters
(Table~\ref{tab:model}), in particular with the mass density coefficient 
$A$ and magnetic braking time scale $\tmb$, translate into an
uncertainty of the predicted X-ray emission 
from stars in the vicinity of Sgr A*. In Fig.~\ref{fig:profile_param},
we demonstrate how the X-ray surface brightness profile is modified by
changing these  parameters within reasonable ranges. We see that with
$\tmb$ fixed at  100~Myr, the data favour (if we were to explain the
entire X-ray emission from Sgr~A* by coronal radiation from spun-up
stars) a rather high stellar density $A\sim$~(4--6)$\times 10^6
M_\odot$~pc$^{-3}$ (taking into account the $\sim 20$\% uncertainty in
the model normalisation due to the conversion to {\sl Chandra}
counts). On the other hand, taking $\tmb=200$~Myr, which is still a
reasonable value for $\lesssim M_\odot$ stars, we can reproduce the
emission from Sgr~A* already with $A\sim 3\times 10^6
M_\odot$~pc$^{-3}$. 
     
\subsection{Approximate scaling formulae}
\label{s:scale}
         
Based on the results of our simulations, we can write down an
approximate formula (which provides good fits to the
  computational results at least up to $A=1.2\times
  10^{7}M_\odot$~pc$^{-3}$) for the expected radial profile of the
  intrinsic, i.e. unabsorbed, X-ray (2--10~keV) luminosity density of
  a stellar cusp of density $\rho_\ast(r)=A(r/0.25~{\rm pc})^{-1.3}$,
  assuming a magnetic braking time $\tmb=100$~Myr: 
\beqa
\ehx(r)\approx 7.5\times 10^{34}A_6^{1.6}
\left(\frac{r}{0.05~\rm{pc}}\right)^{-2}
\nonumber\\
+ 3\times 10^{34}A_6 \left(\frac{r}{0.05~\rm{pc}}\right)^{-1.3}
~{\rm erg}~{\rm s}^{-1}~{\rm pc}^{-3},
\label{eq:ehx_fit}
\eeqa
where $A_6=A/(10^{6}M_\odot~{\rm pc}^{-3})$. 

The first term on the right-hand side of equation~(\ref{eq:ehx_fit})
reflects the contribution of spun-up stars, whereas the second term is
due to the combined emission of recently born (less than a few $\tmb$
ago) and hence rapidly rotating stars and ABs. The latter contibution 
obviously scales linearly with the cusp density. In contrast, the
cumulative X-ray luminosity of spun-up stars grows approximately as
$A^{1.6}$ with increasing cusp density, since it is not only the total
number of stars that increases, but also the relative fraction of fast
rotators among them. The amplitude of the second term should be
regarded as indicative only, because it can vary by a large factor
depending on the actual star formation history in the cusp and the
actual presence of ABs there.

Given the GC distance of 8~kpc and taking into account the
absorption column $\NH\approx 1.1\times 10^{23}$~cm$^{-2}$ towards
Sgr~A* (see Table~\ref{tab:fits} below), equation~(\ref{eq:ehx_fit})
translates into an approximate relation for the expected surface
brightness of a GC stellar cusp in the 2--10~keV energy band: 
\beqa
S_{\rm HX,obs}(\theta)\approx 1.13\times 10^{-15} A_6^{1.6}
\theta^{-1}
\nonumber\\
+ 8.6\times 10^{-16} A_6
\theta^{-0.3}~{\rm erg}~{\rm s}^{-1}~{\rm cm}^{-2}~{\rm arcsec}^{-2},
\label{eq:brightness_fit}
\eeqa
where $\theta$ is the projected distance from Sgr~A* in arcsec. For
$A_6\gtrsim 2$, the first term dominates at $\theta\lesssim$1--2''
from Sgr~A*. 

It is important to emphasise that if magnetic braking, for some reason,
operates less efficiently in spun-up stars in the GC region than in stars
in the solar neighbourhood, i.e. $\tmb>100$~Myr, then rapidly rotating
stars will constitute a larger fraction of all stars in the cusp and
thus produce more X-ray emission per unit stellar mass, e.g. by $\sim
50$\% if $\tmb=200$~Myr for a given value of $A$. A similar effect 
may result if the X-ray luminosity of the stellar cusp is 
dominated by (presumably) much more numerous K/M dwarfs rather than by
Sun-like stars. In either case, it is possible that a somewhat lighter stellar
cluster, than predicted by equation~(\ref{eq:brightness_fit}), could
produce a given X-ray flux from Sgr~A*.
 
\subsection{X-ray spectrum}
\label{s:spec}

We have demonstrated that a high-density cusp of late-type MS stars 
in the GC region can produce X-ray emission with the radial surface
brightness profile as observed by {\sl Chandra} near  
Sgr A*. Will this emission from rapidly spinning stars also have a
spectrum similar to that of Sgr A*?

To answer this question, we re-measured the spectrum of Sgr A*
reported by B03. Since the GC region is rich in high-energy
phenomena, a significant or even dominant fraction of the
X-ray emission observed outside of the innermost few arcseconds 
(see the image in Fig.~\ref{fig:image}) may have a different nature from the 
central X-ray cusp. We therefore constructed a spectrum from the photons
detected within 3 pixels ($\approx 1.5''$) of Sgr A*. We modelled
the detector background for {\sl Chandra}/{\sl ACIS} using the $acisbg$ task
(http://cxc.cfa.harvard.edu/contrib/maxim/acisbg/) based on the stowed
dataset. The background-subtracted 2--8~keV flux from the studied
region is $1.45\times 10^{-13}$~erg~s$^{-1}$~cm$^{-1}$ (with a
statistical uncertainty of a few per cent), in good agreement with B03.  

\begin{figure}
\centering
\includegraphics[bb=60 180 580 700, width=\columnwidth]{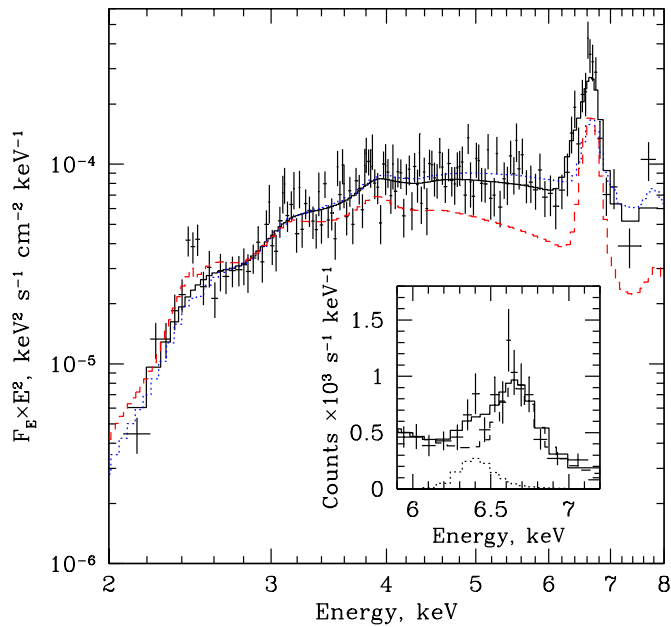}  
\caption{Spectrum of X-ray emission from the region with radius of
  1.5'' around Sgr~A* measured by {\sl Chandra} in the 2--8~keV 
  energy band (data points) and its best fit by the
  $wabs*(cevmkl+gauss)$ model with the parameters given in
  Table~\ref{tab:fits}. For comparison shown are the best-fit $cevmkl$
  models (see Table~\ref{tab:fits}) of {\sl ASCA} spectra of the coronally
  active stars V711~Tau (blue dotted) and 47~Cas~B (red dashed), modified by
  cold gas absorption with $\NH=1.08\times 10^{23}$~cm$^{-2}$ and
  normalised to match the spectral flux of Sgr A* at 3~keV. The inset
  shows a blow-up of the Sgr~A* spectrum at $\sim $6--7~keV. The
  6.4~keV line and CEVMKL components are shown by the dotted and
  dashed lines, respectively.
} 
\label{fig:spectra}
\end{figure}

In agreement with B03, the measured spectrum of Sgr A*
(Fig.~\ref{fig:spectra}) is consistent with thermal emission of
optically thin, hot plasma, modified at low energies by 
absorption in cold gas. In addition, we detect a $\sim 3\sigma$
significant excess that is consisent with a 6.4~keV fluorescent line of neutral
iron. We fitted the {\sl Chandra} data in the 2--8~keV energy band by
different models in XSPEC (version 12.5.1, \citealt{arnaud96}) and found
that a good fit is provided by the model $wabs*(cevmkl+gauss)$. The
best-fit parameters are given in Table~\ref{tab:fits} and the spectral
fit is shown in Fig.~\ref{fig:spectra}. CEVMKL is a sum of MEKAL 
\citep{mewetal85,lieetal95} models for optically thin plasmas with a 
range of temperatures and a power-law distribution of emission
measures, $d{\rm EM}\propto (T/T_{\rm max})^{\alpha-1} dT/T_{\rm
  max}$, where we fix $\alpha=1$ (the results are only weakly
sensitive to this parameter). In fact, using a 
single-temperature MEKAL model instead of CEVMKL provides a similarly
good fit to the strongly absorbed spectrum of Sgr~A*, but we have
chosen CEVMKL because it is more physically justified in
application to X-ray spectra of coronally active stars.

\begin{table*}
\caption{Best-fit spectral parameters for the X-ray emission within 1.5'' 
of Sgr A* and for two coronally active stars. 
\label{tab:fits}
}
\smallskip

\begin{tabular}{cccc}
\hline
Parameter & Sgr A* & V711 Tau & 47 Cas B \\
 & {{\sl Chandra/ACIS-I} (2--8~keV)} & \multicolumn{2}{c}{{\sl ASCA/GIS}
  (2--9~keV)} \\ 
 & $wabs*(cevmkl+gauss)$ & \multicolumn{2}{c}{$cevmkl$} \\
\hline
$T_{\rm max}$ (keV) & $3.2\pm 0.3$ & $3.74\pm 0.08$ & $2.4\pm 0.3$\\
$\alpha$ & 1 (fixed) & 1 (fixed) & 1 (fixed)\\
$A_{\rm Fe}$ & $0.80\pm 0.12$ & $0.32\pm 0.07$ & 1 (fixed)\\
$\NH$, cm$^{-2}$ & $(10.8\pm 0.6)\times 10^{22}$ & --- & --- \\
${\rm EW}_{6.4~{\rm keV}}$ (eV) & $110_{-40}^{+30}$ & --- & --- \\
$\chi^2$/dof & 156/130 & 419/315 & 33/49\\
Unabsorbed 2--10~keV flux (erg~s$^{-1}$~cm$^{-2}$) & $(4.3\pm
0.8)\times 10^{-13}$ & $2.9\times 10^{-11}$ & $2.4\times 10^{-12}$\\
Distance (pc) & $8\times 10^3$ & 29.0 & 33.6 \\
Unabsorbed 2--10~keV luminosity (erg~s$^{-1}$) & 
$(3.3\pm 0.6)\times10^{33}$ & $2.9\times 10^{30}$ & $3.2\times 10^{29}$\\ 
\hline 
\end{tabular}

Notes: The abundances of all elements except Fe were fixed
at solar values \citep{gresau98}. The derived Fe abundance for
V711~Tau is indicative only, given the simplicity of the adopted model.
The quoted distances to the stars are based on {\sl Hipparcos}
parallax measurements \citep{peretal97}. 

\end{table*}

Table~\ref{tab:fits} also gives the absorption-corrected flux
and luminosity of Sgr~A* in the 2--10~keV band for the adopted
spectral model. These values have $\sim 20$\% uncertainties.  
In summary, all of the characteristics of the X-ray spectrum of Sgr~A* 
derived here are in good agreement with the estimates of B03 
(although these authors used a single-temperature thermal
model), but the associated statistical uncertainties have now become
small due to the greatly increased observational time and statistics.

\subsubsection{Comparison with stellar spectra}

According to the reference mass density model (equation~\ref{eq:rho}, 
$A=4\times 10^6 M_\odot$~pc$^{-3}$), which we used to describe the 
X-ray surface brightness profile (Fig.~\ref{fig:profile}), the total
mass of stars and stellar remnants within the cylinder with a
projected radius of 1.5'' around Sgr~A* is $8.4\times 10^4 M_\odot$
($6.3\times 10^4 M_\odot$) within $\pm 0.25$~pc ($\pm 0.1$~pc) along
the line of sight. Since within $\sim 0.1$~pc from Sgr~A* the bulk of
the coronal X-ray luminosity is produced by $\sim 10$\% of low-mass
stars (see the bottom panel of Fig.~\ref{fig:source} and the related
discussion in \S\ref{s:results}) and we assumed $\flms=0.5$, the X-ray flux
accumulated by {\sl Chandra} within 1.5'' of Sgr~A* can be produced by some 3--5
thousand rapidly spinning stars, each emitting 
$10^{29}$--$10^{30}$~erg~s$^{-1}$ in the 2--10~keV energy band.

As was mentioned in \S\ref{s:xray}, there is a well-known
trend of X-ray spectral hardening with increasing coronal luminosity,
which in turn is closely linked to stellar rotation. Although the
rotational velocities are expected to vary from one star to another in
a GC stellar cusp, its cumulative X-ray emission will be dominated  
by stars that are in ($\Omega>\omegac$) or close to being in 
($\Omega\lesssim\omegac$) saturation regime of coronal activity (see
again the bottom panel of Fig.~2). 

An ultimate spectral test of our model would consist in constucting a 
composite X-ray spectrum of an ensemble of stars rotating at various 
angular velocities representing the distribution of $\Omega$ values 
predicted by the model. We feel that performing such a computation is 
premature given the relative simplicity of our current model. In 
particular, this model deals with only one type of star, namely $1 M_\odot$ 
dwarfs. In reality, depending on the composition of the GC stellar cusp, 
its X-ray luminosity may be dominated by K/M dwarfs rather than by less 
numerous Sun-like stars. The rotational velocity distributions and X-ray spectra
may be somewhat different for stars of different types, and a composite
spectrum of the stellar cusp should take this into account. Perhaps
more importantly, the combined spectrum of the stellar cusp may
contain a significant contribution from giant stellar flares (with
$\lx\sim\lbol$), especially at energies above $\sim 3$~keV. As
discussed in \S\ref{s:var} below, the statistics of such flares
remains very limited. Moreover, such events are usually detected by
all-sky monitors and their X-ray spectra are poorly studied at
brightness peak. This renders it practically impossible to estimate
the contribution of giant flares to the cumulative X-ray spectrum of
the stellar cusp. In view of these complications and uncertainties, we
restrict our discussion below to a fairly qualitative comparison of
the spectrum of the extended emission from Sgr A* with the spectra of
a couple of well-studied coronally active stars in the Solar neighbourhood. 

The first object, V711~Tau, is a well-known RS CVn
system. It consists of K0-2~IV and G0-5~V stars, tidally 
locked with a period of 2.84~days \citep{fekel83}. Given the masses and 
radii of the components of $M_1=1.4\pm 0.2 M_\odot$, $M_2=1.1\pm
0.2M_\odot$, $R_1=3.9\pm 0.2R_\odot$ and $R_2=1.3\pm 0.2 R_\odot$
\citep{fekel83}, the subgiant component rotates at a significant
fraction of its breakup velocity, $\Omega/\omegab\approx 0.27$, and is
therefore definitely in saturated regime of coronal activity.
We emphasise that although V711~Tau is an active binary, its spectral
properties should resemble those of a single star that has just been
spun up by a strong tidal interaction to $\Omega>\omegac$. 

The other object, 47~Cas~B, is a relatively young ($\sim 100$~Myr old) 
solar analog (spectral type G0-2~V) belonging to the Pleiades moving 
group \citep{teletal05}. It has a likely rotation period of 
$\sim 1$~day (see \citealt{teletal05} and references therein), so that
$\Omega/\omegab\sim 0.1$, with this fast rotation being due to the
star's young age rather than due to its binarity (the star is 
actually a member of a wide binary system). It might thus be 
close to but not necessarily fully in saturation regime of coronal 
activity. Therefore, 47~Cas~B probably more closely resembles a 
typical spun-up star in the GC stellar cusp than V711~Tau. 

We obtained the X-ray spectra of these two systems using
archival data of the GIS instrument aboard the {\sl ASCA}
observatory. Both spectra are well fit in the 2--9~keV band by the
CEVMKL model. The best-fit parameters are given in
Table~\ref{tab:fits}. The fit for V711~Tau could be improved further 
by allowing the abundances of S, Si and other metals to vary, but such
small modifications are unimportant for our current purpose of 
comparing the spectral continua of Sgr~A* and coronally active 
stars. Fig.~\ref{fig:spectra} shows the best-fit models for V711~Tau 
and 47~Cas~B, modified by cold gas absorption with 
$\NH=1.08\times 10^{23}$~cm$^{-2}$, as measured for
Sgr~A*, and normalised to the flux of Sgr A* at 3~keV.

We see that the spectra of V711~Tau and 47~Cas~B are slightly harder
and somewhat softer, respectively, than the spectrum of Sgr~A*. 
This appears to be consistent with i) V711~Tau having saturated
coronal activity -- as witnessed by its very high hard X-ray
luminosity, $\lhx\sim 3\times 10^{30}$~erg~s$^{-1}$ (2--10~keV,
Table~\ref{tab:fits}) and ii) 47~Cas~B being not quite saturated --
its X-ray luminosity of $\sim 2.5\times 10^{30}$~erg~s$^{-1}$
(0.1--10~keV, \citealt{teletal05}, of which $\sim 3\times
10^{29}$~erg~s$^{-1}$ is at 2--10~keV, Table~\ref{tab:fits}) might be
somewhat below the saturation level of $\sim 10^{-3}\lbol$.

Although it is difficult to draw firm conclusions from this exemplary 
analysis, it indicates that the non-flaring emission of an ensemble of 
spun-up stars can produce a spectrum that is sufficiently hard to 
explain the low-energy (below $\sim 4$~keV) spectral continuum
observed from Sgr A*. However, to account for the hard X-ray (at
energies above $\sim 4$~keV) flux from Sgr A* within the stellar cusp model,
one possibly needs an additional significant contribution from giant
stellar flares. This appears plausible (see \S\ref{s:var} and
\S\ref{s:hard}) but requires further study. 

\subsubsection{Iron abundance}
\label{s:abund}
 
The {\sl Chandra} spectrum of Sgr~A*, in particular the strength of the
6.7-7.0~keV blend of FeXXV and FeXXVI lines, indicates a somewhat
sub-solar ($\sim 0.7$--0.9 relative to the solar value of
\citealt{gresau98}) abundance of iron in the X-ray emitting
plasma. This result appears to be consistent with the model of
cumulative emission of coronally active stars.

Indeed, both V711~Tau and 47~Cas~B exhibit significantly subsolar iron
abundances. For V711~Tau, this is already suggested by the {\sl ASCA} spectrum
presented here: $A_{\rm Fe}\sim 0.3$. More reliable estimates
are provided by analyses of high-resolution spectra obtained with the
Reflection Grating Spectrometer on {\sl XMM-Newton}, which 
indicate $A_{\rm Fe}\sim 0.2$ \citep{audetal03} and $\sim 0.5$
\citep{teletal05} for V711~Tau and 47~Cas~B, respectively (for the Fe
abundance of \citealt{gresau98}). These examples reflect the well-known 
phenomenon of reduced abundance of iron and other low first ionisation
potential elements in active stellar coronae (see \citealt{guedel04}
for a review).

\subsubsection{6.4 keV emission}
\label{s:6.4}

Combined radiation of coronally active stars can also possibly account for 
at least part of the flux of 6.4~keV emission apparent in the {\sl Chandra} 
spectrum of Sgr~A* (see Fig.~\ref{fig:spectra}). Indeed, a
significant fraction (up to 50\%) of the hard X-ray emission 
produced in the corona of a star irradiates the underlying
chromosphere and photosphere and produces fluorescent emission, mainly
in the 6.4~keV line of iron. Such fluorescent emission provides a
means of probing the geometry of coronal flares and the photospheric
abundance of iron.

This problem has been of interest for a long time in application to
the Sun, first theoretically \citep{bai79,basko79} and then
observationally as powerful 6.4~keV emission was actually observed
during solar flares (e.g. \citealt{culetal81,phietal94}). Recently,
6.4~keV emission has also been detected during coronal flares of few
other stars, including the RS CVn binary II~Peg \citep{ostetal07}, the 
G-type giant HR~9024 \citep{tesetal08} and the M~dwarf EV Lac 
\citep{ostetal10}. The measured equivalent widths of the 6.4~keV line
range from a few tens to $\sim 200$~eV, which is similar to the
values reported for solar flares and agrees with the expected flux of
fluorescent emission arising due to irradiation of a stellar
atmosphere of solar Fe abundance by hard X rays from a coronal loop
located at $\lesssim 0.3R_\ast$ above the atmosphere \citep{draetal08}. 

To our knowledge, there have been no reports of 6.4~keV
emission in non-flaring spectra of coronally active stars, probably due
to the insufficient sensitivity and/or low spectral resolution of the
existing instruments and the presence of the much stronger 6.7~keV and
7.0~keV line complexes of iron in coronal spectra. However, based
on the existing evidence for solar and stellar flares we may expect
the integrated spectrum of a cusp of rapidly spinning stars to contain a
6.4~keV line with an equivalent width ${\rm EW}\sim 50$~eV. Moreover, as is
further discussed below in \S\ref{s:fluor}, if giant coronal flares
contribute significantly to the cumulative X-ray emission of the cusp,
somewhat stronger fluorescent emission can be expected, possibly consistent
with the value ${\rm EW}=110^{+30}_{-40}$~eV measured in the spectrum of Sgr~A*.

Another possibility to explain part of the observed 6.4~keV emission
is fluorescent emission from a cold gas permeating the stellar
cusp. However, in order to account for, say, ${\rm EW}\sim 50$~eV, this gas
must have a column density $\sim 3\times 10^{23}$~cm$^{-2}$, assuming
irradiation by $\sim 3.2$~keV thermal emission from Sgr~A*, solar
abundance of iron (${\rm Fe}/{\rm H}=3.16\times 10^{-5}$) and adopting
the ionisation cross sections from \cite{veryak95} and fluorescent
yield of 0.34. This is much larger than the total absorption column of
$\approx 10^{23}$~cm$^{-2}$ measured by {\sl Chandra} towards Sgr~A*. 

We conclude that photospheric fluorescent emission accompanying the
coronal activity of spun-up stars can possibly explain the
tentatively detected 6.4~keV emission from Sgr~A*. 

\section{Discussion}
\label{s:discuss}

We have demonstrated that the combined coronal radiation of  
a high-density cluster of late-type MS stars ``rejuvenated'' by tidal
interactions can in principle explain the main properties of the
quiescent X-ray emission from Sgr~A*, namely its characteristic size (${\rm
  FWHM}\sim 1.5''$), luminosity ($\sim 10^{33}$~erg~s$^{-1}$, observed
in the 2--10~keV band) and possibly (this issue needs further study) 
spectrum (optically thin thermal emission with $kT\lesssim 3.5$~keV, 
absorbed by $\NH\sim 10^{23}$~cm$^{-2}$). In this scenario, extended
X-ray emission is produced by several thousand rapidly spinning stars
(with periods of $\gtrsim 1$~day) with nearly saturated coronal activity. 

Below, we first discuss the implications of  
our model for the mass density of the GC stellar cusp (\S\ref{s:mass}), 
then compare the observational properties of the model with those of truly
diffuse X-ray emission from a hot accretion flow onto the SMBH 
(\S\ref{s:compare}) and finally discuss the implications for the accretion 
onto the SMBH (\S\ref{s:smbh}).

\subsection{Implications for the GC stellar cusp}
\label{s:mass}

The existence of a cusp in the number density distribution of low-mass MS
stars near Sgr~A* has not yet been proven by direct
observations. According to our baseline model, to explain the bulk of
the X-ray emission from Sgr~A* there must be $\sim$~(2--3)$\times
10^4$ Sun-like stars within a column of 1'' radius centered on Sgr~A*
(within 0.25~pc along the line of sight). This number corresponds to a
range $\sim$~(4--6)$\times 10^6M_\odot$~pc$^{-3}$ for the coefficient
$A$ characterising the radial profile of stellar mass density
(equation~\ref{eq:rho}) and to a relative fraction $\flms=0.5$ of
low-mass stars per $M_\odot$. 

Equations~(\ref{eq:ehx_fit}) and (\ref{eq:brightness_fit})    
can be used to predict the contribution of coronal radiation
of low-mass stars (not only spun-up stars but also young stars and active
binaries) to the observed X-ray emission from the Sgr~A* region for a
given density of the GC stellar cusp, assuming that tidally spun-up
stars experience efficient magnetic braking ($\tmb=100$~Myr). 

As follows from our discussion in \S\ref{s:cusp}, the space
density of stars required to explain the quiescent X-ray flux from
Sgr~A* somewhat exceeds the range $A\sim $~(1--3)~$\times 
10^6M_\odot$~pc$^{-3}$, suggested by the available statistics of
bright stars in the central parsec of the
Galaxy. Furthermore, it is possible that low-mass stars near Sgr~A*
are underabundant near Sgr~A*, as suggested by the paucity
of red giants in the central several arcseconds. However, the number
of low-mass stars in the central arcsecond required to explain the
{\sl Chandra} data is consistent with the upper limit deduced from the
measured surface brightness of NIR diffuse light near Sgr~A* (see
equations~\ref{eq:n_diff} and \ref{eq:a_diff}).

Furthermore, it is possible that a cusp of low-mass stars will produce
a higher X-ray luminosity for a given density than implied by our
baseline model:

\begin{enumerate}

\item
First, if magnetic braking operates somewhat less
efficiently than assumed, then a larger fraction of the stars will retain their
fast rotation since their last spin-up episode. As we discussed in
\S\ref{s:scale}, increasing the characteristic time $\tmb$ from
100~Myr to a still reasonable (especially for K/M dwarfs) 200~Myr
leads to a 50\% increase in the luminosity of the stellar
cusp. 

\item
Secondly, as discussed in \S\ref{s:var} below, multiple giant
stellar flares may significantly increase the X-ray luminosity  
of the stellar cusp. 

\item
Thirdly, in our baseline model stellar remnants played a
relatively unimportant role, as we assumed their relative fraction to
be $\fsr=0.25$ and the average mass $\msr=1M_\odot$. In reality, the
innermost region of the Galaxy may be overpopulated by stellar-mass
black holes (e.g. \citealt{moretal93,freetal06,loeetal10}). These heavy objects
can spin up low-mass stars much more efficiently than stars or white
dwarfs (see AK01). 

\item
Finally, in our simulations we implicitly assumed
that stars were in circular orbits around the SMBH. A more accurate 
computation should take into account mutual interactions of stars
moving in eccentric orbits around the SMBH, which might increase the
number of spun-up stars.

\end{enumerate}

In summary, given the remaining uncertainties on
the observational side and the fact that the presented model is
somewhat simplistic, it seems plausible that a factor of few less stars
in the central region might still produce the bulk of the X-ray flux
observed from Sgr~A*. 

\subsection{Comparison with accretion flow models}
\label{s:compare}

The crucial question is how to distinguish observationally
the combined coronal radiation of a cluster of spun-up stars from the diffuse 
emission of a quasi-spherical accretion flow of hot gas onto the
central SMBH, which has been the favoured explanation for the
quiescent X-ray emission from Sgr~A* so far. Below, we discuss
differences and similarities between the two models.

\subsubsection{Variability and flares}
\label{s:var}

We have so far assumed that the X-ray luminosity emitted by a
coronally active star is solely determined by its current rotation
rate. In reality, virtually all stars with coronae experience 
flares during which their X-ray luminosity increases by orders of
magnitude. Therefore, the cumulative X-ray emission of a GC stellar cusp
must be variable at some level.

Statistics of coronal flares in stars of different types implies (see
\S13 in the review of \citealt{guedel04} and references therein) that 
X-ray/UV flare luminosities (or total energies) obey a power
law with a slope of $\approx 2$, i.e. $dN/dL\propto L^{-2}$, and it
remains unclear if the strongest or weakest flares dominate in the
long term-integrated energy output of stellar coronae and if the
"quiescent" luminosity is in fact just a superposition of weak flares. 

Particularly interesting are giant flares observed from
nearby RS CVn systems. Only a few dozen such flares have been
detected in several decades of X-ray astronomical observations by
all-sky monitors (see \citealt{areetal03}). In particular, the SSI
instrument aboard the {\sl Ariel V} spacecraft detected 7 flares that 
were attributed to RS CVn binaries \citep{pyemch83}, including 2
flares from one of the best studied systems II~Peg. Interestingly, all
these flares had similar properties, namely peak luminosities of (1--6)$\times
10^{32}$~erg~s$^{-1}$ (in the 2--18~keV energy band), characteristic
durations $\sim 5$~hours and total emitted energies (in the X-ray
band) $\sim 6\times 10^{36}$~erg~s$^{-1}$. Based on this statistics,
\citet{pyemch83} concluded that a typical RS CVn binary produces about
1 such giant flare per year. This explains why dedicated observations of
individual RS CVn systems by X-ray telescopes, which 
typically last less than a day, fail to detect such extreme flares.

On Dec. 16, 2005, another giant flare was detected from II~Peg by the 
hard X-ray instrument BAT and X-ray telescope (XRT) aboard the
{\sl Swift} spacecraft \citep{ostetal07}. These observations provided 
uniquely detailed, broad-band information on a giant flare. During the
flare, the luminosity peaked at $\sim 1.3\times 10^{33}$ and $\sim 8\times
10^{32}$~erg~s$^{-1}$ in the 0.8--10~keV and 10--200~keV,
respectively. Thus, the total X-ray luminosity of the corona for 
1--2~hours reached $\gtrsim 40$\% of the bolometric luminosity of the
binary system ($\sim 5\times 10^{33}$~erg~s$^{-1}$,
\citealt{mewetal97}). For comparison, the non-flaring X-ray
(0.5--12~keV) luminosity of II~Peg is $\sim 10^{31}$~erg~s$^{-1}$
\citep{hueetal01}, or $\sim10^{-3}$ of its bolometric
luminosity. This implies that stellar coronae that are already in a 
saturation regime can brighten up by an additional factor of $\gtrsim
100$ during giant flares.

New data on giant flares are now coming from the MAXI all-sky monitor
on the {\sl International Space Station}. Already during the first
year of operation it detected 14 flares from 7 RS CVn binaries,
including 2 huge flares with peak (1.5--10~keV) luminosities $\sim
5\times 10^{33}$~erg~s$^{-1}$ from II~Peg and GT~Mus
\citep{tsuetal11}. These new results, first, confirm that giant 
flares occur roughly once a year per RS CVn system and, second, suggest that 
stellar coronae may produce even more extreme flares than was 
considered possible before.

The spun-up stars in a GC cusp should be similar to RS CVn systems as
regards their coronal activity, apart from being less luminous by a
factor of few, since the X-ray bright component in RS CVn binaries is
often a subgiant rather than a normal star. We may thus expect every
spun-up star to undergo a giant X-ray flare with a luminosity of
$10^{32}$--$10^{33}$~erg~s$^{-1}$ and duration of a few hours roughly
once a year. A "superflare" of the M dwarf EV Lac \citep{ostetal10},
already mentioned in \S\ref{s:6.4}, at the peak of which the X-ray
(0.3--100~keV) luminosity reached more than 3 bolometric luminosities
of the star, might serve as a good example of such a giant flare on a
single coronally active star (note, however, that this particular
flare was fairly short, $\sim 0.5$~hour).

Given that some 3--5 thousand spun-up stars are required to
produce the measured X-ray flux from Sgr~A*, we may expect several
stars to experience giant flares at any moment. The cumulative 
luminosity of these flares may be comparable or exceed the total
luminosity of the remaining several thousand "quiescent" stars. In
fact, due to flaring activity the integrated X-ray luminosity of a GC
stellar cusp can be significantly higher than expected based on 
equations~(\ref{eq:lx}) and (\ref{eq:lhx}).

If the cumulative emission of spun-up stars is dominated by strong
flares, we can expect the X-ray flux from Sgr~A* to vary on 
hourly and daily time scales by tens of per cent or even by a factor
of few, due to the changing number and intensity of flares. This
variability should be especially strong at higher energies,
i.e. around 10~keV and above, because the coronal emission spectrum
hardens during flares, as for example was the case
during the superflare of II~Peg \citep{ostetal07}. 
In principle, the already existing {\sl Chandra} and {\sl XMM-Newton}
data can be used to verify this prediction. Furthermore, it is possible
that some of the weaker, $\sim 1$~hour long X-ray flares
observed from Sgr~A* (e.g., \citealt{bagetal03,poretal08}) were in
fact produced by stellar coronae. 

At the same time, it is clear that the stronger X-ray flares observed 
from Sgr~A*, during which its luminosity reaches several $\times 10^{34}$ 
and even more than $10^{35}$~erg~s$^{-1}$, are of different origin 
and most likely produced in the close vicinity of the SMBH (see 
\citealt{genetal10} for review). In particular, by analysing the {\sl Chandra} 
data for some of such flares we have verified that their origin is consistent
with a point source (within the {\sl Chandra} PSF) at the position of Sgr A*. 

\subsubsection{Multifrequency properties}
\label{s:multi}

The primary radiative component of active stellar coronae is optically thin
X-ray emission of a multitemperature gas, with a dominant temperature 
$kT\sim$~a few keV. A similar X-ray spectrum can be produced by a hot
accretion flow if the accretion rate is sufficiently low. In this
case, the bulk of the total X-ray luminosity is due to gas near the 
Bondi radius $\rb\sim 0.05$~pc, whereas the additional emission
produced in the inner regions of the presumably advection-dominated
accretion flow is relatively unimportant. As argued by
\citet{quataert02}, such a situation may be realised in the quiescent
state of the SMBH associated with Sgr~A*, for which $\lhx\sim
10^{-11}\ledd$, where $\lhx$ and $\ledd$ are the observed X-ray
luminosity and the SMBH Eddington luminosity, respectively.

Given the same thermal character of the X-ray emission in the scenarios
of a cluster of spun-up stars and hot accretion flow onto the SMBH, are
there spectral features that could distinguish these models from one
another?

\paragraph{Thermal X-ray lines}
\label{s:thermal_lines}

Both models naturally produce coronal X-ray lines, in particular 
a 6.7~keV line of FeXXV. As we showed in \S\ref{s:abund},
the {\sl Chandra} spectrum in the 6.7~keV region indicates a somewhat
subsolar abundance of Fe for Sgr~A*, which is in line with the known
trend for active stellar coronae. However, the measured deviation from
the solar abundance is minor ($A_{\rm Fe}\approx 0.7$--0.9) and does
not seem to contradict the hot accretion flow model for Sgr~A* 
either. Indeed, recent studies of chemical abundances in the
atmospheres of supergiants belonging to young stellar clusters near
Sgr~A* have demonstrated that these several million-old stars were
formed from an interstellar medium with a nearly solar Fe/H ratio (see
\citealt{davetal09} and references therein). Therefore, one can expect the gas
currently feeding the SMBH to have approximately solar Fe abundance as well.

What could the widths of the X-ray lines tell us? In the stellar cusp
model, the line widths should reflect the emission-weighted, radially 
integrated distribution of projected stellar velocities. 
Given equation~(\ref{eq:sigma}) and the model surface brightness profile 
shown in Fig.~\ref{fig:profile}, the lines produced 
within $\sim 1''$ of Sgr~A* should be broadened by $\sigma\sim
500$~km~s$^{-1}$, which corresponds to ${\rm FWHM}\sim 25$--30~eV for the 6.7
line of iron. Hence, the lines are unresolvable with CCD detectors. 

It is more difficult to predict the line widths in the case of a hot 
accretion flow onto the SMBH. In ADAF-type models, ions tend
to acquire the virial velocities in the gravitational potential of the
black hole \citep{naryi94}. However, in the case of Sgr~A* most of the
luminosity measured by {\sl Chandra} from the central arcsecond is
believed to be produced near the Bondi radius. In
this case, the broadening of the lines of heavy elements is probably 
determined by bulk motions in the gas, which should be of the order
of the stellar velocity dispersion if the gas is supplied by winds of
hot stars (see, e.g., \citealt{quataert04,cuaetal06}). 

Therefore, to a first approximation, thermal X-ray lines should be
similarly broadened in both scenarios. 

\paragraph{6.4~keV line}
\label{s:fluor}

A stand-alone issue is 6.4~keV emission of neutral (or weakly ionised) iron. 
As we discussed in \S\ref{s:6.4}, we have tentatively detected such a
line, with an equivalent width ${\rm EW}=110_{-40}^{+30}$~eV in the
{\sl Chandra} spectrum. 

There is no obvious reason to expect 6.4~keV emission in the scenario
of a hot accretion flow onto the SMBH. On the contrary, as we discussed
in \S\ref{s:6.4}, a 6.4~keV line is naturally produced in stellar coronae,
due to the irradiation of the underlying photosphere by the coronal radiation
and subsequent fluorscence. We noted, however, that "quiescent" coronae
are unlikely to produce 6.4~keV emission stronger than ${\rm EW}\sim 50$~eV.
However, during strong flares, the coronal spectrum becomes
significantly harder (see below), hence a larger fraction of the total
luminosity is emitted above the ionisation thereshold of iron
(7.1~keV) and consequently a 6.4~keV line with a larger equivalent width can be
produced. For example, during a ``superflare'' from the M~dwarf
EV~Lac,  6.4~keV emission with ${\rm EW}\sim 200$~eV was detected by
{\sl Swift}/XRT \citep{ostetal10}.

Therefore, if strong flares significantly contribute to the X-ray
luminosity of the stellar cusp, we may expect a strong 6.4~keV line
with ${\rm EW}\lesssim 100$~eV to be present in the X-ray
spectrum. This feature, if confirmed with higher statistical
significance by future observations (in particular, by long 
observations with the {\sl Chandra} High Energy Tranmission Grating,
HETG), is thus a clear indicator in favour of the stellar cusp model 
for the extended X-ray emission from Sgr~A*.

\paragraph{Hard X-ray emission}
\label{s:hard}

If the X-ray source at Sgr~A* is associated with a Bondi/ADAF
accretion flow onto the SMBH, then in addition to the dominant soft
X-ray emission originating in the vicinity of the Bondi radius 
there should be a hard X-ray continuum due to the bremsstrahlung emission
from the much hotter accretion flow at $R\ll\rb$, whose luminosity (at
energies above 10~keV) might be $\sim 1/3$ of the X-ray
luminosity (below 10~keV) of Sgr~A* \citep{quataert02}. Non-thermal
(e.g. synchrotron or syncrotron-self-Compton) emission produced in the
innermost region of the accretion flow may provide some additional
hard X-ray flux. Therefore, a hard X-ray continuum is a generic
feature of Bondi/ADAF models for Sgr~A*.

A significant hard X-ray continuum is also expected in the
case of a cluster of spun-up, coronally active stars. Indeed, as we have
already discussed, strong coronal flares can contribute significantly
to the total X-ray luminosity of the GC cusp. It is well-known (see
e.g. \citealt{densch89,aschwanden02} for reviews) that in solar flares
only part of the total energy released via magnetic reconnection goes
into heating of thermal plasma, and only a fraction of this thermal
energy is radiated away as X-ray (and UV) emission. A significant
amount of energy goes into acceleration of non-thermal electrons and
ions. The non-thermal electrons lose a small fraction of their energy
as hard X-ray bremsstrahlung and radio gyrosynchrotron emission. 

During solar flares, non-thermal hard X-ray emission is some 5 orders 
of magnitude weaker than thermal X-ray emission (see Fig.~94 
in \citealt{aschwanden02}). However, during the 2005 giant flare of
the RS CVn binary II~Peg, its hard X-ray and soft X-ray luminosities
were comparable (see S\ref{s:var} above). Therefore, if strong flares
play a significant role in the combined emission of the GC stellar cusp, 
it might emit in total $\lesssim 10^{33}$~erg~s$^{-1}$ in hard X-rays
(at energies above 10~keV), corresponding to a hard X-ray flux
$\lesssim 10^{-13}$~erg~s$^{-1}$~cm$^{-2}$ for the GC distance of $\approx
8$~kpc. Such a signal can in principle be detected by the {\sl NuSTAR}
telescope, to be launched in 2012, which will have $\sim 10''$ (FWHM)
angular resolution and reach sensitivity $\sim
10^{-14}$~erg~s$^{-1}$~cm$^{-2}$ in the 10--30~keV energy band for an 
exposure of 1~Ms \citep{haretal10}. As we discussed in
\S\ref{s:var} above, this hard X-ray signal should be significantly
variable on time scales of hours and days as a result of flaring
activity on stars. 

We conclude that both the stellar cusp and the accretion flow models
predict hard X-ray bremsstrahlung emission at some significant
level, although these predictions are very uncertain at the moment. 

\paragraph{Gamma-ray emission}

Besides electrons, also protons and ions are accelerated in coronal
flares, reaching kinetic energies up to GEVs. In the case of the Sun,
just a small fraction of such sub-cosmic rays escape into
interplanetary space, while the majority eventually 
hit the lower corona and chromosphere and lose energy there (see 
\citealt{densch89,aschwanden02} for reviews). A small fraction of the
energy lost by these particles (and also by high-energy electrons) is
emitted as gamma-ray continuum and gamma-ray lines.

Similarly to the case of hard X-ray emission, it is difficult to
predict the gamma-ray luminosity of a cluster of spun-up stars. In
solar flares, comparable amounts of energy are emitted in the hard
X-ray and gamma-ray bands, which reflects the more fundamental fact that
similar amounts of energy are injected into acceleration of electrons
and ions (see \S7 and in particular Figs.~86 and 94 in
\citealt{aschwanden02}). Therefore, following our suggestion that a GC
stellar cusp may produce up to $\sim 10^{33}$~erg~s$^{-1}$ in hard X
rays, it is possible that a similar luminosity is also emitted in 
gamma-rays, mainly at 0.5--10~MeV. 

This predicted flux of gamma quanta, $\lesssim 10^{-7}({\rm
  MeV}/E)$~s$^{-1}$~cm$^{-2}$ (where $E$ is the photon energy), is too
weak to be detected by present-day gamma-ray telescopes. For example,
the SPI spectrometer aboard the {\sl INTEGRAL} observatory, after
observing the Galactic bulge region for more than 10~Ms, detects
511~keV positron-annihilation emission with a flux $\sim
10^{-3}$~s$^{-1}$~cm$^{-2}$ and provides upper limits $\sim
10^{-4}$~s$^{-1}$~cm$^{-2}$ on gamma-ray emission in the 0.511--10~MeV
energy range \citep{chuetal11}. Moreover, the angular resolution of
the existing gamma-ray instruments is by far insufficient to study the
Sgr~A* region. Therefore, detection of gamma-ray emission from a GC
stellar cusp may be a matter of distant future.

As concerns the scenario of a hot accretion flow onto the SMBH,
\citet{mahetal97} attempted to predict the gamma-ray luminosity of 
Sgr~A* in the ADAF context. Some of their models resulted in a strong
flux $\sim 10^{-7}$~s$^{-1}$~cm$^{-2}$ of $\sim 100$~MeV photons due
to the decay of neutral pions, with a much weaker emission being
produced at both lower and higher energies. However, we feel that
these estimates should be regarded as indicative at best due to the
many assumptions involved.  

\paragraph{Radio emission}

Related to the production of hard X rays and gamma-rays is the
question of radio emission. The radio emission of stellar coronae has
been well studied, so we can make relatively robust estimates in
this case. It has been shown \citep{gueben93} that across several
decades in luminosity there is a linear correlation between the soft
X-ray and radio (6 or 3.6~cm) luminosities, $\lx/L_{\rm R}\sim
10^{5.5}$, the same for different classes of stars. Moreover, roughly the same
relation seems to hold for large coronal flares. Indeed,
\citet{ostetal07} discuss that during giant flares of RS CVn binaries 
such as II~Peg, V711~Tau and UX~Ari, the radio luminosity exceeded the
quiescent levels by a factor of $\gtrsim 100$
\citep{waletal03,ricetal03}, i.e. by nearly the same factor as the
X-ray luminosity. 

Given that the radio spectra of coronally active stars are approximately 
flat in the centimeter--decimeter range (e.g. \citealt{garetal03}), and
using the X-ray--radio correlation mentioned above, we may estimate
the radio luminosity of a GC stellar cusp at $\nu L_\nu\lesssim
10^{28}$~erg~s$^{-1}$. For comparison, the measured quiescent
luminosity of Sgr~A* is much higher, namely $\sim 4\times 10^{32}$,
$\sim 5\times 10^{31}$ and $\sim 6\times 10^{30}$~erg~s$^{-1}$ at 3.6,
26 and 90~cm, respectively \citep{anetal05}.

We conclude that a GC stellar cusp is unlikely to contribute
significantly to the radio emission of Sgr~A* and thus to affect the
generally accepted view that this emission is produced in the inner
accretion flow onto the SMBH. 

We summarise our comparison of the stellar cusp and hot 
accretion flow models in Table~\ref{tab:models}, where we 
indicate which of the observable properties of Sgr A* can possibly
be accounted for by either model. We see that some properties 
(namely the iron 6.4 keV line) favour the stellar model, others 
(radio emission and strong flares of Sgr A*) are only consistent with 
the hot accretion flow model, and some (hard X-ray emission) allow
both possibilities at the moment. This suggests that coronal radiation
from spun-up stars may dominate in the extended X-ray emission around
Sgr A*, whereas the hot accretion flow onto the SMBH possibly
contributes significantly to the harder X-ray component and is also
resposible for the phemonena spatially associated with the close
vicinity of Sgr A*, such as the radiosource itself and strong flaring activity.

\begin{table}
\caption{Consistency of stellar cusp and hot accretion flow
    models with available data.
\label{tab:models}
}
\smallskip

\begin{tabular}{ccc}
\hline
\multicolumn{1}{c}{Component} &
\multicolumn{1}{c}{Stellar cusp} &
\multicolumn{1}{c}{Accretion flow} \\
\hline
\multicolumn{3}{c}{Quiescent emission} \\
X-ray radial profile & yes & yes \\  
X-ray luminosity & yes & yes \\
Soft X-ray spectrum & yes & yes \\
Hard X-ray spectrum & ? & ? \\
Fe 6.4 keV line & yes & no \\
Radio source & no & yes \\
Faraday rotation limit & yes & ? \\
\multicolumn{3}{c}{Variable emission} \\
Strong flares & no & yes \\
Weak flares & yes & yes \\
\hline
\end{tabular}
\end{table}

\subsection{Implications for gas accretion onto the SMBH}
\label{s:smbh}

If a major fraction of the extended X-ray emission from Sgr~A* is
produced by a high-density stellar cluster, it will have a number of
implications for the growth of the central SMBH, as considered below.

\subsubsection{Accretion rate}
\label{s:bondi}

Previously, the accretion rate through the outer boundary of the
Bondi-type accretion flow onto the SMBH was estimated at $\mdout\sim
10^{-5}M_\odot$~yr$^{-1}$ (e.g. \citealt{yuaetal03}), based 
on the density and temperature of the presumably diffuse gas
at $\rb\sim 0.05$~pc, as measured with {\sl Chandra}. As we already 
discussed in \S\ref{s:intro}, this accretion rate in combination with the very
low ($\lesssim 10^{36}$~erg~s$^{-1}$) bolometric luminosity of Sgr A*
puts a very strong upper limit on the radiative efficiency of accretion
onto the SMBH and hence strong requirements for ADAF-type 
models for Sgr A*. On the other hand, Faraday rotation measurements 
in the direction of Sgr~A* imply that accretion onto the SMBH occurs at 
$\mdin\lesssim 10^{-7}M_\odot$~yr~$^{-1}$ \citep{bowetal03}. 

To satisfy these observational constraints, so-called radiatively
inefficient accretion flow models were suggested for Sgr~A*. In
these models, in contrast to the original ADAF concept, very little
mass available at large radii actually accretes onto the black hole,
while most of the gas outflows or circulates in convective motions
(see \citealt{yuaetal03} and references therein). Hence, in this
scenario, the SMBH accretes gas at a rate $\mdin\lesssim
10^{-7}M_\odot$~yr$^{-1}\ll\mdout $, which allows one to both obtain
the necessary X-ray luminosity and to satisfy the rotation measure limit.

In our scenario, if the bulk of the quiescent X-ray emission from Sgr~A*
is produced by spun-up stars, then the external accretion rate,
$\mdout$, itself may be significantly lower than believed before. In addition,
since the stars are now responsible for the bulk of the X-ray luminosity of
Sgr~A*, the internal accretion rate, $\mdin$, will be lower than
thought before. This should be taken into account and can actually 
be helpful in future developments of hot accretion flow models 
for Sgr~A*.  

\subsubsection{Gas supply by stars}
\label{s:gas}

On the other hand, the spun-up stars themselves can provide gas for 
accretion onto the SMBH via stellar winds. Although mass loss
from solar-like stars increases with increasing coronal activity,
this trend reverses near the saturation limit of coronal activity, so
that the maximal loss rate of a~few~$10^{-12} M_\odot$~yr$^{-1}$ is
experienced by $\sim 1$~Gyr old stars (in the solar neighbourhood), i.e. by
moderately fast rotators \citep{wooetal05}. Therefore, since the total
mass of stars within 0.05~pc of the SMBH is at most a~few~$10^4
M_\odot$, the integrated mass loss rate by stellar winds from the
late-type MS stars cannot exceed $10^{-7}M_\odot$~yr$^{-1}$. However,
as follows from our preceeding discussion is \S\ref{s:bondi}, even
with this low injection rate stellar winds may contribute
significantly to the total accretion rate of the SMBH. 

A potentially more interesting mechanism of gas supply is tidal
interactions of stars. Indeed, SPH simulations suggest that
collision-like encounters, namely those with $\rp\sim R_\ast$, 
not only cause a strong spin-up but also a significant loss of mass from
the star(s), $\sim 0.1 M_\odot$ per collision (AK01,
\citealt{freben05}). For the mass density profile
(equation~\ref{eq:rho}) with $A\sim 3\times 10^6 M_\odot$~pc$^{-3}$, such
close encounters will occur every $\sim 3\times 10^4$ ($\sim 2\times 10^4$) 
years within 0.02 (0.1) pc of Sgr~A* (most collisions take place in
the innermost 0.02~pc). Since these encounters occur well within the
SMBH sphere of influence, we can expect that every few $10^4$~years
$\lesssim 0.1 M_\odot$ of stellar debris will accrete onto 
the SMBH and produce an outburst. Less frequently, even larger amounts of
gas ($\lesssim 1 M_\odot$) can be provided by catastrophic collisions of stars,
i.e. those leading to their complete distruction, and by close
collisions of stars with neutron stars and stellar-mass black holes, which are
likely to be present in significant numbers in the central stellar cluster.

The injection of $\sim 0.1$--$1M_\odot$ of gas from a stellar collision
onto the SMBH can cause an outburst at a significant fraction of its 
Eddington lumonisity. Indeed, the free-fall time to the SMBH from a
typical stellar collision distance of $\sim 0.02$~pc is $\sim
20$~years. Since the infalling gas will need to get rid of
its angular momentum on its way to the black hole, the accretion
episode will probably last $\sim$~$10^{2}$~years. This implies an
accretion rate of $\sim 10^{-3}$--$10^{-2}M_\odot$~yr$^{-1}$ (if all of the gas
actually reaches the black hole), or $\sim 10^{-2}$--$10^{-1}$ of the critical
accretion rate for the $\approx 4\times 10^6M_\odot$ SMBH. Therefore,
stellar collisions can cause outbursts of Sgr~A* with luminosity 
$\lesssim 10^{42\div43}$~erg~s$^{-1}$ every $\sim
10^4$--$10^5$~years. Even stronger but shorter outbursts are expected to occur 
at a similar rate due to tidal distruction of stars in the loss cone
by the SMBH (see \citealt{alexander05} for review). 

The detection of reflected X-ray emission from the Sgr B2
  molecular cloud in the GC region implies that Sgr A* experienced an outburst 
  with a luminosity of few $\times 10^{39}$~erg~s$^{-1}$ (2--200~keV)
  that lasted at least $\sim 10$~years and ended some 100--300~years ago
  \citep{sunetal93,revetal04,teretal10}. Such an event could possibly be caused
  by accretion onto the SMBH of gas torn off from a star that
  experienced a close collision with another star or stellar
  remnant in the nuclear cluster. Moreover, the data also allow for 
  the possibility that this outburst lasted longer than 10~years and
  was stronger at its outset. In such a case, it might have been
  caused by a catastrophic collision of stars.
We also note that \citet{crasun02} searched for traces of X-ray
outbursts of Sgr~A* in the past, looking for delayed X-ray emission
reflected from the neutral atomic and molecular gas distributed over
the Milky Way. They concluded that there was no prolonged X-ray
activity of Sgr A* at or above one per cent of its Eddington level
(i.e. $\sim 5\times 10^{42}$~erg~s$^{-1}$) in the last $\sim
80,000$~years. This limit is consistent with the above estimates for
tidal interaction and destruction of stars.

\section{Conclusion}
\label{s:concl}

We have demonstrated that a significant or even dominant fraction of 
the extended X-ray emission from Sgr A* can be produced by a putative
cusp of low-mass MS stars. To account for the bulk of the extended
emission, there must be $\sim$~(2--3)$\times 10^4$ Sun-like stars
within a column of 1'' radius centered on Sgr~A*, 
of which roughly every ten's star is predicted to be rapidly rotating 
and producing strong thermal X-ray emission with $kT\lesssim
$~a~few~keV as a result of tidal spin-ups caused by close encounters
with other stars and stellar remnants. Turning the argument around,
the {\sl Chandra} data place an interesting upper limit on the space
density of (currently unobservable) low-mass MS stars near
Sgr~A*. This bound is close to and consistent with current constraints
on the central stellar cusp provided by NIR observations.

The model is based on  well-understood physical mechanisms, namely
tidal spin-up due to close fly-bys of stars and stellar 
remnants, generation of coronal activity by stellar rotation and
braking of stellar rotation by magnetised wind.

We have discussed potentially observable signatures of the stellar
cusp model in comparison with the scenario of a hot accretion 
flow onto the SMBH. We demonstrated that the cumulative luminosity 
of spun-up stars in the central cusp can be variable by tens of
per cent or even by a factor of few on hourly and daily timescales as a
result of giant flares occuring on various stars. This variability
could be searched for in archival {\sl Chandra} and {\sl
  XMM-Newton} data. Moreover, it is possible that some of the weaker
($\lhx\sim$~a few~$\times 10^{33}$~erg~s$^{-1}$), $\sim 1$~hour long
X-ray flares from Sgr~A* were produced by stellar coronae. There is
also a chance to detect variable coronal emission from the GC stellar
cusp at energies above 10~keV with the upcoming {\sl NuSTAR} mission.

We have tentatively detected a 6.4~keV line with an equivalent width
of $110^{+30}_{-40}$~eV in the {\sl Chandra} spectrum of Sgr~A*. We
argued that such a line can be expected in the stellar cusp model as a
result of irradiation of the stellar photospheres by the hard X-ray coronal
radiation. On the other hand, such emission is not expected to be
produced in a hot accretion flow onto the SMBH. It is thus very
important to verify the presence of a 6.4~keV line in the Sgr~A*
spectrum with more data. The planned ultradeep {\sl Chandra}
HETG observations of Sgr A* (PI: Baganoff) could solve this task.

On the other hand, the stellar cusp model cannot account for the radio
emission and strong multiwavelength flares of Sgr A*. These are likely
associated with the inner regions of a hot accretion flow onto the
SMBH. Moreover, it is not yet clear if an ensemble of spun-up stars
can explain the hard X-ray emission (at energies above $\sim 4$~keV)
as easily as the softer emission from Sgr A*, since the spectra of
typical coronally active stars may be insufficiently hard. We have
suggested that giant stellar flares might provide a significant
contribution to the total X-ray flux and thus significantly harden the
cumulative spectrum of the stellar cusp, but this issue needs further study.  

Finally, it is plausible that spun-up stars make up the bulk of the 
softer X-ray emission from Sgr A* while a hot accretion flow 
contributes significantly at higher energies.

\section*{Acknowledgments}
We thank the referee for helpful comments and suggestions. 
The research made use of grants RFBR 09-02-00867a, RFBR 10-02-00492a and 
NSh-5069.2010.2, programs the Russian Academy of Sciences P-19 and
OFN-16 and the President's grant MD-1832.2011.2. SS and MR acknowledge
the support of the Dynasty Foundation. 



\begin{thebibliography}{99}

\bibitem[\protect\citeauthoryear{Alexander}{1999}]{alexander99} 
Alexander T., 1999, ApJ, 527, 835 

\bibitem[\protect\citeauthoryear{Alexander}{2005}]{alexander05} 
Alexander T., 2005, PhR, 419, 65 

\bibitem[\protect\citeauthoryear{Alexander 
\& Kumar}{2001}]{alekum01} Alexander T., Kumar P., 2001, ApJ, 549,
  948 (AK01)

\bibitem[\protect\citeauthoryear{An et al.}{2005}]{anetal05} An 
T., Goss W.~M., Zhao J.-H., Hong X.~Y., Roy S., Rao A.~P., Shen Z.-Q., 
2005, ApJ, 634, L49 

\bibitem[\protect\citeauthoryear{Arefiev, Priedhorsky, 
\& Borozdin}{2003}]{areetal03} Arefiev V.~A., Priedhorsky W.~C.,
  Borozdin K.~N., 2003, ApJ, 586, 1238  

\bibitem[\protect\citeauthoryear{Arnaud}{1996}]{arnaud96} Arnaud
K.~A., 1996, ASPC, 101, 17

\bibitem[\protect\citeauthoryear{Aschwanden}{2002}]{aschwanden02} 
Aschwanden M.~J., 2002, SSRv, 101, 1 

\bibitem[\protect\citeauthoryear{Audard et 
al.}{2003}]{audetal03} Audard M., G{\"u}del M., Sres A., Raassen
  A.~J.~J., Mewe R., 2003, A\&A, 398, 1137  

\bibitem[\protect\citeauthoryear{Baganoff et 
al.}{2003}]{bagetal03} Baganoff F.~K., et al., 2003, ApJ, 591, 
891 (B03)

\bibitem[\protect\citeauthoryear{Bahcall 
\& Wolf}{1977}]{bahwol77} Bahcall J.~N., Wolf R.~A., 1977, ApJ, 216,
  883 

\bibitem[\protect\citeauthoryear{Bai}{1979}]{bai79} Bai T., 
1979, SoPh, 62, 113 

\bibitem[\protect\citeauthoryear{Bartko et al.}{2010}]{baretal10} 
Bartko H., et al., 2010, ApJ, 708, 834 

\bibitem[\protect\citeauthoryear{Basko}{1979}]{basko79} Basko 
M.~M., 1979, SvA, 23, 224 

\bibitem[\protect\citeauthoryear{Binney 
\& Tremaine}{1987}]{bintre87} Binney J., Tremaine S., 1987, in Galactic
  Dynamics (Princeton: Princeton Univ. Press), 541

\bibitem[\protect\citeauthoryear{Bower et al.}{2003}]{bowetal03} 
Bower G.~C., Wright M.~C.~H., Falcke H., Backer D.~C., 2003, ApJ, 588, 331 

\bibitem[\protect\citeauthoryear{Buchholz, Sch{\"o}del, 
\& Eckart}{2009}]{bucetal09} Buchholz R.~M., Sch{\"o}del R., Eckart
  A., 2009, A\&A, 499, 483

\bibitem[\protect\citeauthoryear{Churazov et 
al.}{2011}]{chuetal11} Churazov E., Sazonov S., Tsygankov S., 
Sunyaev R., Varshalovich D., 2011, MNRAS, 411, 1727 

\bibitem[\protect\citeauthoryear{Cramphorn 
\& Sunyaev}{2002}]{crasun02} Cramphorn C.~K., Sunyaev R.~A., 2002,
  A\&A, 389, 252  

\bibitem[\protect\citeauthoryear{Cuadra et al.}{2006}]{cuaetal06} 
Cuadra J., Nayakshin S., Springel V., Di Matteo T., 2006, MNRAS, 366, 358 

\bibitem[\protect\citeauthoryear{Culhane et 
al.}{1981}]{culetal81} Culhane J.~L., et al., 1981, ApJ, 244, 
L141 

\bibitem[\protect\citeauthoryear{Dale et al.}{2009}]{daletal09} 
Dale J.~E., Davies M.~B., Church R.~P., Freitag M., 2009, MNRAS, 393,
1016 

\bibitem[\protect\citeauthoryear{Davies et al.}{2009}]{davetal09} 
Davies B., Origlia L., Kudritzki R.-P., Figer D.~F., Rich R.~M., Najarro 
F., 2009, ApJ, 694, 46 

\bibitem[\protect\citeauthoryear{Dennis 
\& Schwartz}{1989}]{densch89} Dennis B.~R., Schwartz R.~A., 1989, SoPh, 121, 75 

\bibitem[\protect\citeauthoryear{Do et al.}{2009}]{doetal09} Do 
T., Ghez A.~M., Morris M.~R., Lu J.~R., Matthews K., Yelda S., Larkin J., 
2009, ApJ, 703, 1323 

\bibitem[\protect\citeauthoryear{Drake, Ercolano, 
\& Swartz}{2008}]{draetal08} Drake J.~J., Ercolano B., Swartz D.~A., 2008, ApJ, 678, 385 

\bibitem[\protect\citeauthoryear{Fekel}{1983}]{fekel83} Fekel 
F.~C., Jr., 1983, ApJ, 268, 274 

\bibitem[\protect\citeauthoryear{Freitag, Amaro-Seoane, 
\& Kalogera}{2006}]{freetal06} Freitag M., Amaro-Seoane P., Kalogera
  V., 2006, ApJ, 649, 91  

\bibitem[\protect\citeauthoryear{Freitag 
\& Benz}{2005}]{freben05} Freitag M., Benz W., 2005, MNRAS, 358, 1133

\bibitem[\protect\citeauthoryear{Garc{\'{\i}}a-S{\'a}nchez, Paredes, 
\& Rib{\'o}}{2003}]{garetal03} Garc{\'{\i}}a-S{\'a}nchez J., Paredes
  J.~M., Rib{\'o} M., 2003, A\&A, 403, 613  

\bibitem[\protect\citeauthoryear{Genzel et al.}{1996}]{genetal96} 
Genzel R., Thatte N., Krabbe A., Kroker H., Tacconi-Garman L.~E., 1996, 
ApJ, 472, 153 

\bibitem[\protect\citeauthoryear{Genzel, Eisenhauer, 
\& Gillessen}{2010}]{genetal10} Genzel R., Eisenhauer F., Gillessen
  S., 2010, RvMP, 82, 3121  
 
\bibitem[\protect\citeauthoryear{Ghez et al.}{2008}]{gheetal08} 
Ghez A.~M., et al., 2008, ApJ, 689, 1044 

\bibitem[\protect\citeauthoryear{Gillessen et 
al.}{2009}]{giletal09} Gillessen S., Eisenhauer F., Trippe S., 
Alexander T., Genzel R., Martins F., Ott T., 2009, ApJ, 692, 1075 

\bibitem[\protect\citeauthoryear{Graham 
\& Spitler}{2009}]{graspi09} Graham A.~W., Spitler L.~R., 2009, MNRAS,
  397, 2148

\bibitem[\protect\citeauthoryear{Grevesse 
\& Sauval}{1998}]{gresau98} Grevesse N., Sauval A.~J., 1998, SSRv, 85,
  161 

\bibitem[\protect\citeauthoryear{Guedel 
\& Benz}{1993}]{gueben93} Guedel M., Benz A.~O., 1993, ApJ, 405, L63 

\bibitem[\protect\citeauthoryear{G{\"u}del}{2004}]{guedel04} G{\"u}del
  M., 2004, A\&ARv, 12, 71  

\bibitem[\protect\citeauthoryear{Harrison et 
al.}{2010}]{haretal10} Harrison F.~A., et al., 2010, SPIE, 7732,  

\bibitem[\protect\citeauthoryear{Huenemoerder, Canizares, 
\& Schulz}{2001}]{hueetal01} Huenemoerder D.~P., Canizares C.~R.,
  Schulz N.~S., 2001, ApJ, 559, 1135  

\bibitem[\protect\citeauthoryear{Liedahl, Osterheld, 
\& Goldstein}{1995}]{lieetal95} Liedahl D.~A., Osterheld A.~L.,
  Goldstein W.~H., 1995, ApJ, 438, L115 

\bibitem[\protect\citeauthoryear{L{\"o}ckmann, Baumgardt, 
\& Kroupa}{2010}]{loeetal10} L{\"o}ckmann U., Baumgardt H., Kroupa P.,
  2010, MNRAS, 402, 519  

\bibitem[\protect\citeauthoryear{Mahadevan, Narayan, 
\& Krolik}{1997}]{mahetal97} Mahadevan R., Narayan R., Krolik J.,
  1997, ApJ, 486, 268

\bibitem[\protect\citeauthoryear{Mewe, Gronenschild, 
\& van den Oord}{1985}]{mewetal85} Mewe R., Gronenschild E.~H.~B.~M., van den Oord G.~H.~J., 1985, A\&AS, 62, 197 

\bibitem[\protect\citeauthoryear{Mewe et 
al.}{1997}]{mewetal97} Mewe R., Kaastra J.~S., van den Oord G.~H.~J.,
  Vink J., Tawara Y., 1997, A\&A, 320, 147  

\bibitem[\protect\citeauthoryear{Morris}{1993}]{moretal93} Morris 
M., 1993, ApJ, 408, 496 

\bibitem[\protect\citeauthoryear{Muno et al.}{2004}]{munetal04} 
Muno M.~P., et al., 2004, ApJ, 613, 326 

\bibitem[\protect\citeauthoryear{Muno et al.}{2008}]{munetal08} 
Muno M.~P., Baganoff F.~K., Brandt W.~N., Morris M.~R., Starck J.-L., 2008, 
ApJ, 673, 251 

\bibitem[\protect\citeauthoryear{Muno et al.}{2009}]{munetal09} 
Muno M.~P., et al., 2009, ApJS, 181, 110 

\bibitem[\protect\citeauthoryear{Narayan 
\& Yi}{1994}]{naryi94} Narayan R., Yi I., 1994, ApJ, 428, L13 

\bibitem[\protect\citeauthoryear{Nayakshin 
\& Sunyaev}{2005}]{naysun05} Nayakshin S., Sunyaev R., 2005, MNRAS, 364, L23 

\bibitem[\protect\citeauthoryear{Osten et al.}{2007}]{ostetal07} 
Osten R.~A., Drake S., Tueller J., Cummings J., Perri M., Moretti A., 
Covino S., 2007, ApJ, 654, 1052 

\bibitem[\protect\citeauthoryear{Osten et al.}{2010}]{ostetal10} 
Osten R.~A., et al., 2010, ApJ, 721, 785 

\bibitem[\protect\citeauthoryear{Paumard et 
al.}{2006}]{pauetal06} Paumard T., et al., 2006, ApJ, 643, 1011 

\bibitem[\protect\citeauthoryear{Perets}{2009}]{perets09} Perets 
H.~B., 2009, ApJ, 690, 795 

\bibitem[\protect\citeauthoryear{Perryman 
\& ESA}{1997}]{peretal97} Perryman M.~A.~C., et al., 1997, The Hipparcos
  and Tycho catalogues, ESASP, 1200   \

\bibitem[\protect\citeauthoryear{Phillips et 
al.}{1994}]{phietal94} Phillips K.~J.~H., Pike C.~D., Lang J., 
Watanbe T., Takahashi M., 1994, ApJ, 435, 888 

\bibitem[\protect\citeauthoryear{Porquet et 
al.}{2008}]{poretal08} Porquet D., et al., 2008, A\&A, 488, 549 

\bibitem[\protect\citeauthoryear{Pye 
\& McHardy}{1983}]{pyemch83} Pye J.~P., McHardy I.~M., 1983, MNRAS, 205, 875 

\bibitem[\protect\citeauthoryear{Quataert}{2002}]{quataert02} 
Quataert E., 2002, ApJ, 575, 855 

\bibitem[\protect\citeauthoryear{Quataert}{2004}]{quataert04} 
Quataert E., 2004, ApJ, 613, 322 

\bibitem[\protect\citeauthoryear{Revnivtsev et 
al.}{2004}]{revetal04} Revnivtsev M.~G., et al., 2004, A\&A, 425, L49 

\bibitem[\protect\citeauthoryear{Revnivtsev et 
al.}{2006}]{revetal06} Revnivtsev M., Sazonov S., Gilfanov M.,
  Churazov E., Sunyaev R., 2006, A\&A, 452, 169  

\bibitem[\protect\citeauthoryear{Revnivtsev, Vikhlinin, 
\& Sazonov}{2007}]{revetal07} Revnivtsev M., Vikhlinin A., Sazonov S.,
  2007, A\&A, 473, 857

\bibitem[\protect\citeauthoryear{Revnivtsev et 
al.}{2009}]{revetal09} Revnivtsev M., Sazonov S., Churazov E., 
Forman W., Vikhlinin A., Sunyaev R., 2009, Natur, 458, 1142 

\bibitem[\protect\citeauthoryear{Richards et 
al.}{2003}]{ricetal03} Richards M.~T., Waltman E.~B., Ghigo 
F.~D., Richards D.~S.~P., 2003, ApJS, 147, 337 

\bibitem[\protect\citeauthoryear{Saar 
\& Brandenburg}{1999}]{saabra99} Saar S.~H., Brandenburg A., 1999,
  ApJ, 524, 295

\bibitem[\protect\citeauthoryear{Sazonov et 
al.}{2006}]{sazetal06} Sazonov S., Revnivtsev M., Gilfanov
  M., Churazov E., Sunyaev R., 2006, A\&A, 450, 117  

\bibitem[\protect\citeauthoryear{Sch{\"o}del et 
al.}{2007}]{schetal07} Sch{\"o}del R., et al., 2007, A\&A, 469, 125 

\bibitem[\protect\citeauthoryear{Sch{\"o}del, Merritt, 
\& Eckart}{2009}]{schetal09} Sch{\"o}del R., Merritt D., Eckart A.,
  2009, A\&A, 502, 91  

\bibitem[\protect\citeauthoryear{Schrijver 
\& Zwaan}{2000}]{schzwa00} Schrijver C.~J., Zwaan C., 2000, Solar and
  Stellar Magnetic Activity. Cambridge Univ. Press, Cambridge 

\bibitem[\protect\citeauthoryear{Skumanich}{1972}]{skumanich72} 
Skumanich A., 1972, ApJ, 171, 565 

\bibitem[\protect\citeauthoryear{Strassmeier et 
al.}{1993}]{stretal93} Strassmeier K.~G., Hall D.~S., Fekel F.~C., Scheck M., 1993, A\&AS, 100, 173 

\bibitem[\protect\citeauthoryear{Sunyaev, Markevitch, 
\& Pavlinsky}{1993}]{sunetal93} Sunyaev R.~A., Markevitch M.,
  Pavlinsky M., 1993, ApJ, 407, 606

\bibitem[\protect\citeauthoryear{Telleschi et 
al.}{2005}]{teletal05} Telleschi A., G{\"u}del M., Briggs K., 
Audard M., Ness J.-U., Skinner S.~L., 2005, ApJ, 622, 653 

\bibitem[\protect\citeauthoryear{Terrier et 
al.}{2010}]{teretal10} Terrier R., et al., 2010, ApJ, 719, 143 

\bibitem[\protect\citeauthoryear{Testa et al.}{2008}]{tesetal08} 
Testa P., Drake J.~J., Ercolano B., Reale F., Huenemoerder D.~P., Affer L., 
Micela G., Garcia-Alvarez D., 2008, ApJ, 675, L97 

\bibitem[\protect\citeauthoryear{Trippe et 
al.}{2008}]{trietal08} Trippe S., et al., 2008, A\&A, 492, 419 

\bibitem[\protect\citeauthoryear{Tsuboi et al.}{2011}]{tsuetal11}
  Tsuboi Y., et al.,, 2011, Proc. 4th international MAXI symposium, in
  press 

\bibitem[\protect\citeauthoryear{Verner 
\& Yakovlev}{1995}]{veryak95} Verner D.~A., Yakovlev D.~G., 1995,
  A\&AS, 109, 125  

\bibitem[\protect\citeauthoryear{Waldram et 
al.}{2003}]{waletal03} Waldram E.~M., Pooley G.~G., Grainge 
K.~J.~B., Jones M.~E., Saunders R.~D.~E., Scott P.~F., Taylor A.~C., 2003, 
MNRAS, 342, 915 

\bibitem[\protect\citeauthoryear{Wang, Dong, 
\& Lang}{2006}]{wanetal06} Wang Q.~D., Dong H., Lang C., 2006, MNRAS, 371, 38 

\bibitem[\protect\citeauthoryear{Weisskopf et 
al.}{2002}]{weietal02} Weisskopf M.~C., Brinkman B., Canizares 
C., Garmire G., Murray S., Van Speybroeck L.~P., 2002, PASP, 114, 1 

\bibitem[\protect\citeauthoryear{Wood et al.}{2005}]{wooetal05} 
Wood B.~E., M{\"u}ller H.-R., Zank G.~P., Linsky J.~L., Redfield S., 2005, 
ApJ, 628, L143 

\bibitem[\protect\citeauthoryear{Xu et al.}{2006}]{xuetal06} Xu 
Y.-D., Narayan R., Quataert E., Yuan F., Baganoff F.~K., 2006, ApJ, 640, 
319 

\bibitem[\protect\citeauthoryear{Yuan, Quataert, 
\& Narayan}{2003}]{yuaetal03} Yuan F., Quataert E., Narayan R., 2003,
  ApJ, 598, 301
  
\bibitem[\protect\citeauthoryear{Yusef-Zadeh et 
al.}{2009}]{yusetal09} Yusef-Zadeh F., et al., 2009, ApJ, 706, 
348 

\bibitem[\protect\citeauthoryear{Yusef-Zadeh et al.}{2011}]{yusetal11}
  Yusef-Zadeh, F., Bushouse, H., Wardle, M., 2011, ApJ (in press),
  arXiv:1109.2175

\end{thebibliography}
\end{document}